\documentclass[twocolumn]{aastex61}
\pdfoutput=1 
\usepackage{amsmath,amstext}
\usepackage[T1]{fontenc}
\usepackage{apjfonts} 
\usepackage[figure,figure*]{hypcap}

\def\alwaysmath#1{\ifmmode{#1}\else{$#1$}\fi} 
\newcommand{\teff}{$T_{\rm{eff}}$}
\newcommand{\logg}{$\log{g}$}
\newcommand{\microt}{$\xi_{\rm{t}}$}

\newcommand{\feh}{[Fe/H]}
\newcommand{\cfe}{[C/Fe]}
\newcommand{\nfe}{[N/Fe]}
\newcommand{\afe}{[$\alpha$/Fe]}

\newcommand{\ctwo}{$\chi^2$}

\newcommand{\sersic}{S\'ersic}
\newcommand{\abs}[1]{\left\lvert#1\right\rvert}
\lefthead{Garc\'{\i}a P\'erez et al.}

\shorttitle{AASTeX 6.1 Template}
\shortauthors{Author A et al.}

\begin{document}

\title{The Bulge Metallicity Distribution from the APOGEE Survey}

\author{Ana E. Garc\'{\i}a P\'erez}
\affiliation{Department of Astronomy, University of Virginia, Charlottesville, VA 22904-4325, USA}
\affiliation{Instituto de Astrof\'{\i}sica de Canarias, E-38205 La Laguna, Tenerife, Spain}
\affiliation{Departamento de Astrof\'{\i}sica, Universidad de La Laguna, E-38206 La Laguna, Tenerife, Spain}
\author{Melissa Ness}
\affiliation{Max-Planck-Institut f\"ur Astronomie, K\"onigstuhl 17, D-69117 Heidelberg, Germany}
\author{Annie C. Robin}
\affiliation{Institut Utinam, CNRS UMR 6213, OSU THETA, Universit\'e Bourgogne-Franche-Comt\'e, 41bis avenue de l'Observatoire, 25000 Besan\c{c}on, France}
\author{Inmaculada Martinez-Valpuesta}
\affiliation{Instituto de Astrof\'{\i}sica de Canarias, E-38205 La Laguna, Tenerife, Spain}
\affiliation{Departamento de Astrof\'{\i}sica, Universidad de La Laguna, E-38206 La Laguna, Tenerife, Spain}
\author{Jennifer Sobeck}
\affiliation{Department of Astronomy, University of Washington, Seattle, WA 98195-1580, USA}
\author{Gail Zasowski}
\affiliation{Department of Physics \& Astronomy, University of Utah, Salt Lake City, UT 84112, USA}
\author{Steven R. Majewski}
\affiliation{Department of Astronomy, University of Virginia, Charlottesville, VA 22904-4325, USA}
\author{Jo Bovy}
\affiliation{Department of Astronomy and Astrophysics, University of Toronto, 50 St. George Street, Toronto, ON, M5S 3H4, Canada}
\affiliation{Dunlap Institute for Astronomy and Astrophysics, University of Toronto, 50 St. George Street, Toronto, Ontario, M5S 3H4, Canada}
\author{Carlos Allende Prieto}
\affiliation{Instituto de Astrof\'{\i}sica de Canarias, E-38205 La Laguna, Tenerife, Spain}
\affiliation{Departamento de Astrof\'{\i}sica, Universidad de La Laguna, E-38206 La Laguna, Tenerife, Spain}
\author{Katia Cunha}
\affiliation{Observat\'orio Nacional, S\~ao Crist\'ov\~ao, Rio de Janeiro, Brazil}
\affiliation{Steward Observatory, University of Arizona, 933 North Cherry Avenue, Tucson, AZ 85721, USA}
\author{L\'eo Girardi}
\affiliation{Laborat\'orio Interinstitucional de e-Astronomia - LIneA, Rua Gal. Jos\'e Cristino 77, Rio de Janeiro, RJ - 20921-400, Brazil}
\affiliation{Osservatorio Astronomico di Padova, INAF, Vicolo dell'Osservatorio 5, I-35122 Padova, Italy}
\author{Szabolcs M\'esz\'aros}
\affiliation{ELTE E\"otv\"os Lor\'and University, Gothard Astrophysical Observatory, Szombathely, Hungary}
\affiliation{Premium Postdoctoral Fellow of the Hungarian Academy of Sciences} 
\author{David Nidever}
\affiliation{National Optical Astronomy Observatories, Tucson, AZ 85719, USA}
\author{Ricardo P. Schiavon}
\affiliation{Astrophysics Research Institute, Liverpool John Moores University, 146 Brownlow Hill, Liverpool, L3 5RF
United Kingdom}
\author{Mathias Schultheis}
\affiliation{Laboratoire Lagrange, Universit\'e C\^ote d'Azur, Observatoire de la C\^ote d'Azur, CNRS, Blvd de l'Observatoire, F-06304 Nice, France}
\author{Matthew Shetrone}
\affiliation{University of Texas at Austin, McDonald Observatory, Fort Davis, TX 79734, USA}
\author{Verne V. Smith}
\affiliation{National Optical Astronomy Observatories, Tucson, AZ 85719, USA}

\begin{abstract}

The Apache Point Observatory Galactic Evolution Experiment (APOGEE) provides spectroscopic information of regions of the inner Milky Way  inaccessible to optical surveys. We present the first large study of the metallicity distribution of the innermost Galactic regions based on homogeneous measurements from the SDSS Data Release 12 for 7545 red giant stars within 4.5\,kpc of the Galactic center, with the goal to shed light on the structure and origin of the Galactic bulge.

Stellar metallicities are found, through multiple-Gaussian decompositions, to be distributed in several components indicative of the presence  of various stellar populations such as the bar, or the thin and the thick disk. A super-solar (\feh$=+0.32$) and a solar (\feh$\ =+0.00$) metallicity components, tentatively associated with the thin disk and the Galactic bar, respectively, seem to be the major contributors near the midplane. A solar-metallicity component extends outwards in the midplane but is not observed in the innermost regions. The central regions (within 3\,kpc of the Galactic center) reveal, on the other hand, the presence of a significant metal-poor population (\feh$\ =-0.46$), tentatively associated with the thick disk, and which becomes the dominant component far from the midplane  ($\abs{Z} \geqslant +0.75$\,kpc). Varying contributions from  these different components produce a transition region at $+0.5$\,\rm{kpc} $ \leqslant \abs{Z} \leqslant +1.0$\,kpc 
characterized by a significant vertical metallicity gradient.
 
\end{abstract}


\section{Introduction}

In the standard theoretical framework for galaxy formation and evolution, galaxy formation proceeds by hierarchical merging of cold dark matter clumps and their 
associated baryons. 
However, the physics that drives the evolution of baryonic matter, critical for realistically modeling 
the luminous components of galaxies, remains
to be understood. Processes such as star formation and feedback work on scales much smaller than 
the resolution of current galaxy simulations, which limits the generation of robust predictions \citep[e.g.,][]{Agertz11}.

The Milky Way (MW) bulge is an exemplar of a barred bulge \citep{Dwek95} 
with a low \sersic\ index \citep{Widrow08} and an X-shaped profile \citep{McWilliam10, Nataf10}. 
N-body simulations of disk galaxies have demonstrated that bar formation and bar instabilities are 
important for the evolution of central regions in spiral galaxies \citep{Combes90, Raha91, Athanassoula05}. 
Bars can form in thin disks and then buckle, which explains observations of rotationally supported bars, 
peanut shapes, and X-shape profiles in the inner regions of galaxies \citep{Bureau05}. 

Simulations of Milky-Way  like galaxies can form bars and reproduce at least some of the observed
properties of the MW bulge  \citep[e.g.,][]{Guedes11, Okamoto13}. However, the direct attribution of 
MW bulge properties to bar instabilities and buckling has not yet being established. A vertical metallicity gradient, 
which has been detected in the MW bulge, was originally thought to be unsustainable after bar buckling due to 
orbital mixing, but as discussed by \cite{Ness13},  is indeed possible \citep[see also][]{Martinez13}.

The bulk of the Milky Way's bulge stellar population is old
\citep[10\,Gyr, e.g.,][]{Ortolani95, Clarkson08}, but observations
of microlensed turnoff dwarfs \citep{Bensby13}, intermediate mass 
asymptotic giant branch stars \citep[AGB,][]{Uttenthaler07} and planetary nebulae \citep[PNe][]{anibal14} 
provide evidence of a younger ( $<$ 5 Gyr)
population \citep[see also][]{Gesicki14}\footnote{A young and metal-rich population is also seen in the inner regions of the Andromeda Galaxy \citep{Boyer13}.}. A wide range of metallicities is observed in the bulge, with the mode around solar.
Kinematical investigation of metal-rich M-type giants by the BRAVA survey
\citep{Rich07, Howard09} revealed that the bulge has cylindrical rotation, 
leaving little room for a hotter kinematical component (a classical bulge) \citep{Shen10}. 
However, the bulge also has distinct sub-populations that hint
at a complex formation history. These multiple populations create metallicity 
gradients, which are not reproduced by disk galaxy simulations ignoring the mixing of populations.

\citet{Zoccali08} observed fields along the minor axis (at $b \leqslant -4\degr$) 
and measured vertical metallicity variations of $-0.5$\,dex\,kpc$^{-1}$. This outer vertical 
gradient was later confirmed by the Gaia-ESO \citep{GAIA12, Rojas14} and ARGOS surveys \citep{Freeman13, Ness13}.  
Using photometric data from the Vista Variables in the Via Lactea (VVV) 
program, \cite{Gonzalez13} created a map, mainly of the Southern bulge, showing a
smooth metallicity variation with Galactic longitude and a flattening of the vertical
gradient in the inner regions ($|b|\lesssim5\degr$). This flattening was first found at high spectral resolution by
\cite{Rich12} in a sample of 44 M-type giants. Observing $\sim$430 stars in the
red clump with a modest resolving power ($R=6,500$), \cite{Babusiaux14} confirmed a flattening in the 
innermost parts, in fields at $b=0$\degr, $l= +10, +6, -6$\degr, and $b=1$\degr, $l=0$\degr).

A weaker longitudinal metallicity gradient is present in the inner bulge region, as seen clearly in 
the metallicity map of \cite{Gonzalez13}. This behavior with Galactic longitude was confirmed at 
higher spectral resolution for $b\sim-3.5\degr$ by the GIRAFFE Inner Bulge Survey (GIBS) in \cite{Gonzalez15}.

These metallicity gradients are not a consequence of a single narrow metallicity
distribution that shifts in mean metallicity as a function of $b$. Instead, these gradients appear to reflect the varying contribution of different populations 
\citep[e.g.,][]{Hill11, Ness13, Rojas14, Gonzalez15}. The different scale-heights of various metallicity 
sub-populations are tied to their different kinematics. 
The most dramatic example of this effect is the X-shaped bulge. Metal-rich 
stars are preferentially associated with this structure \citep[e.g.,][]{Hill11, Ness13, Rojas14, Zasowski16}, while  
metal poor stars are not (e.g., \citealt{Uttenthaler12, Ness13}). This association may 
be explained by the way stars are redistributed as a function of their initial birth radius into 
the bulge \citep{diMatteo14}. Some studies assign the metal-poor stars to a spherical component 
(e.g., \citealt{Hill11, Dekany13, Rojas14, Zoccali17}), while others associate them with a disk-like 
structure (e.g., \citealt{Ness16, Portail17}). 

Many of the observed properties of the bulge described above
have been reproduced in recent cosmological simulations of galaxy
formation \citep{Okamoto13, Martig12, Inoue12}.  Simulations are now 
capable of following the evolution of baryons throughout the history of the 
Universe, and therefore can model long-timescale secular processes, such 
as the formation of a younger population of stars in the inner galaxy as the result
of gas flows driven by internal dynamical formation processes \citep{Obreja13,Ness14}. 

R12While the recent successes of cosmological models are encouraging, 
such models rely on simple recipes for handling the sub-grid physics and initial 
conditions. The improvement of these models 
can only be accomplished through increasingly detailed observations, which permit the 
refinement of both the initial conditions and the sub-grid physics.

The MW is an invaluable tool in addressing the complicated problem of correctly simulating 
spiral galaxies as it is possible to resolve its constituent stars into separate subpopulations.  Quantitative 
knowledge about vertical and radial metallicity gradients in the MW bulge, particularly at low Galactic latitude, 
are key.  Parameterizing the metallicity gradients in detail across the bulge into the disk, and from the 
midplane outwards to high latitude, is critical to understanding the bulge's formation history and 
ultimately being able to produce self consistent simulations capable of describing the large scale 
properties of our Galaxy. 

The Apache Point Observatory Galactic Evolution Experiment (APOGEE; \citealt{Majewski17}), 
a program of the Sloan Digital Sky Survey III \citep[SDSS-III;][]{Eisenstein11}, has produced the most complete chemokinematical 
database of stars useful for mapping the properties of the inner Galaxy based on high quality spectra at a resolution of $R=22,500$.  
The APOGEE $H$-band wavelength observations easily penetrate the heavily dust-extincted portions of the MW bulge and disk,
and therefore they allow for the study of the \feh\ variations not only in the outer bulge but---importantly---in  
poorly-studied low-latitude regions, including the inner bulge and along the Galactic plane. 
As the APOGEE survey has covered predominantly the northern part of the bulge, it complements Gaia-ESO, BRAVA, ARGOS and GIBS, which are primarily Southern Hemisphere surveys.  In addition, APOGEE has provided more detailed chemical information, 
with individual element abundances for around 15 atomic species. This has made possible the identification of chemically peculiar groups of stars in the Galactic bulge \citep{aegp13,Schiavon17}.

Kinematical and metallicity 2D map ($l$, $b$) based on APOGEE DR12 data have been presented in \cite{Ness16}. 
They show a bulge  rotating cylindrically and with small gradients of radial velocities and metallicities in the innermost regions.  
The present work focuses on the 
{\it metallicity distribution function} (MDF) of the inner Galaxy and its 3D variations.  
We pay careful attention to possible biases in the APOGEE DR12 results, and how they might influence the APOGEE 
mapping of bulge chemistry.  The structure of the paper is as follows: 
Section\,\ref{obsmet} provides a  
description of the observations and summarizes how APOGEE determines stellar metallicities, 
an assessment of sample selection effects is given in
Section\,\ref{bias}, while Section\,\ref{samdist}
addresses how distances are determined and used to
winnow the sample to stars in the central Galaxy. 
Metallicity maps in Galactic Cartesian coordinates with origin at the Galactic center 
($X$, $Y$, $Z$) and the distribution of individual metallicities are presented in Section\,\ref{map}. The APOGEE results are discussed in terms of bulge structure models in Section\,\ref{modcomp} and final conclusions are offered in Section\,\ref{consec}.

\begin{figure}
\figurenum{1}
\begin{center}
\includegraphics[trim=0.5cm 0cm 2cm 15cm, clip,scale=0.5]{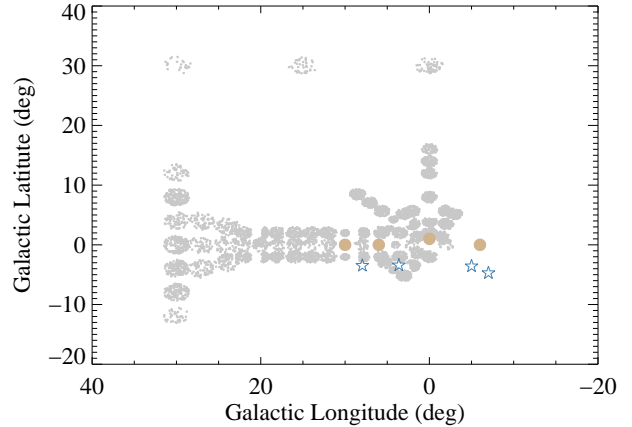}
\caption{Distribution of the 7545 stars within 4.5\,kpc from the Galactic Center included in this study.  Fields with $b > 40\degr$ are not shown for clarity. Each displayed field is {2-3\degr} in diameter and contains typically $\sim 100$ stars. The GIBS fields observed at high spectral resolution (stars) and \citet[][circles]{Babusiaux14} fields are marked.}
\end{center}

\label{maps}
\end{figure}

\section{Observations \& Metallicities}
\label{obsmet}
\subsection{Observations}
\label{obs}

The results presented in this paper are for a sample of 7545 giant stars observed in the APOGEE 
survey between July 2011 and July 2014, and that have
estimated distances that place them within 4.5\,kpc from the Galactic Center (GC, $d_\mathrm{GC} \leqslant 4.5$\,kpc). This choice was motivated by sampling the edges of the long-bar \citep{Wegg15}. The stars are distributed in 83 APOGEE pointings of 
typically 100 stars each, 
as illustrated in Figure\,\ref{maps}.  The individual circular fields, ranging from 1 to 3 degrees 
in diameter, are identified by their Galactic coordinates ($l$, $b$) in degrees, e.g., 000+02.
Five fields toward the central Galaxy --- the APOGEE field centered on the Galactic Center (000+00), 
a BRAVA field (000$-$05), Baade's Window (001$-$04), 
and the Sagittarius fields, SGRC and SGRCM-4 --- were excluded due to  
their special target selection criteria. 
The observations cover both the outer bulge ($|l|$ or $|b| > 4\degr$) and 
importantly, the poorly studied low latitude bulge ($|b| < 4\degr$). 
Sixteen of the eighty three fields lie in the inner bulge, $|(l,\ b)| \leqslant 4\degr$, while
twenty nine outer fields 
(up to $l=30\degr$) have $|b|\leqslant2\degr$. 
Previously such low-latitude regions had very few stars observed with high-resolution spectroscopy: only a few dozen M-type giants with 
$|b| \leqslant 2\degr$ and a few hundred G-type and K-type giants. The vast majority of the sample is at distances
between 4 and 12 kpc from the Sun, and suffers line-of-sight extinctions between 0.2 and 1 magnitudes.

The APOGEE $H$-band spectra, 
acquired with a cryogenically~-cooled, multi-object spectrograph \citep{Wilson12} coupled to the
Sloan Foundation 2.5m telescope \citep{Gunn06} at Apache Point Observatory, in New Mexico, and recorded by 
three HAWAII-2RG detectors, span the wavelength range 1.51--1.69\,${\mu}m$. The instrument has 300 input fibers and,  
in the standard APOGEE configuration,  approximately 230 fibers are assigned to science 
targets\footnote{The number of stars per field analyzed here is further reduced by our quality criteria 
and fiber-to-fiber distance limitations (see Sec. 3 and 4).}  and 70 are reserved for calibration: 
35 targeting hot stars to record the telluric absorption pattern plus 35 sky fibers. 
To achieve Nyquist sampling at the shortest wavelengths, 
multiple exposures are taken while dithering the detector array by 
half a pixel \citep[see][for further details]{Nidever15}.

For the bulge fields, stars were selected from the 2MASS Point Source Catalog \citep{Skrutskie06} 
by color and magnitude, largely adopting $[J-K_{\rm{s}}]_0\geqslant0.5$ 
and $7\leqslant H \leqslant 11$, although 
a few fields were considered part of the APOGEE disk sample and slightly different selection criteria, with deeper integrations, 
were applied for them. The specified color cut was adopted to minimize the contamination by foreground dwarf stars 
(most prominent at $[J-K_{\mathrm{s}}]_0 < 0.8$), retaining potential low metallicity giants in the sample.  
The faint limit of $H < 11$ was set to ensure a minimum signal-to-noise ratio (S/N) of 100 per pixel for 
any inner Galaxy fields where single, approximately 1-hour visits were used \citep[as opposed to the 3-visit norm 
for APOGEE fields; for more details see][]{Zasowski13}.  Only $\sim687$ of the $7545$ stars presented here 
have $S/N < 100$ and all have $S/N\geqslant50$. Reddening corrections were estimated by combining near- and 
mid-IR photometry (2MASS, IRAC, and WISE), using the RJCE method \citep{Majewski11} and the \citet{Indebetouw05} 
extinction law.

Raw data were processed with APOGEE's custom data reduction pipeline \citep{Nidever15}, 
following a standard procedure: pixel dither combination, spectral extraction, wavelength calibration, 
sky emission and telluric contamination correction,
and (when applicable) visit combination. All the spectra have been publicly released as part of the SDSS Data Release 12 \citep[DR12,][]{Alam15}.


 \subsection{Metallicity Determination and Sample Selection}
 \label{metal}
 
The APOGEE Stellar Parameter and Abundance Pipeline (ASPCAP; \citealt{ASPCAP}) was employed to 
determine stellar metallicities (\feh) 
simultaneously with the other atmospheric parameters \teff, \logg, \cfe, \nfe, and {\afe}.  
ASPCAP relies upon {\ctwo}  minimization to match each star's entire APOGEE spectrum to a library of 
pre-computed, LSF-convolved (FWHM resolving power $R\equiv \lambda/\delta\lambda \sim22,500$), 
and normalized synthetic spectra \citep{Shetrone15, Zamora15}.  The microturbulence was tied to the 
surface gravity value by the relation \microt = $2.478-0.325\log g$, derived from the analysis of a sub-sample of 
APOGEE data. The final ASPCAP metallicities were calibrated to well-known values of a sample of 
globular and open cluster stars \citep{Holtzman15}.  Based on the dispersion around the calibration 
values, the typical metallicity accuracy is estimated to be about 0.12 dex. However, the 
precision of the measurements is significantly better, typically about 0.05 dex (Holtzman et al. 2015).  
This precision is usually enhanced when  abundance ratios such as [O/Fe] are considered. 
In fact, Nidever et al. (2014) found, for red clump stars in
the thin disk, a spread in [$\alpha$/Fe]  at any given [Fe/H] between $0.02$-$0.04$\,dex, 
for high signal-to-noise ratios, and Bertr\'an de Lis et al. (2016) found that stars in open clusters 
with similar temperatures showed consistent [O/Fe] ratios to within 0.01 dex. 

Our sample is dominated by cooler red giants, and 
consequently, our metallicities are slightly more uncertain than the bulk of the APOGEE sample, 
around $0.05-0.09$ dex. Additional  details about the APOGEE DR12 
extracted parameters and abundances may be found in \citep{Holtzman15}. 

To select stars with reliable ASPCAP parameters, the APOGEE\_ASPCAPFLAG bitmask flag 
was used.\footnote{https://www.sdss3.org/dr12/algorithms/bitmasks.php} For the present study, 
stars were removed from our sample if any of the following were true: unreliable {\teff}, {\logg}, 
or {\feh}; large differences between photometric and spectroscopic {\teff} estimates; 
large {\ctwo} values; low $S/N$; and indications of rapid rotation from broadened line profiles.  
The final sample stars have parameters inside the ranges $-0.4\lesssim\ $\logg\ $\lesssim 4.0$, $3600 \leqslant$ {\teff} $\ \leqslant 5500$\,K, 
and $-2.7 \lesssim $ \feh\ $\lesssim +0.6$, which match those where the DR12 atmospheric 
parameters are calibrated.  Known cluster members, as reflected in the APOGEE\_TARGET1 and APOGEE\_TARGET2 flags, 
were removed from the sample. In addition, stars with a radial 
velocity dispersion ($vscatter$) larger than $1$\,km\,s$^{-1}$ were excluded, since that is usually 
an indication of binarity.



\section{Metallicity Bias}
\label{bias}

\begin{figure}
\figurenum{2}
\begin{center}
\includegraphics[angle=0,trim=2.5cm 2cm 15cm 5cm, clip,scale=0.7]{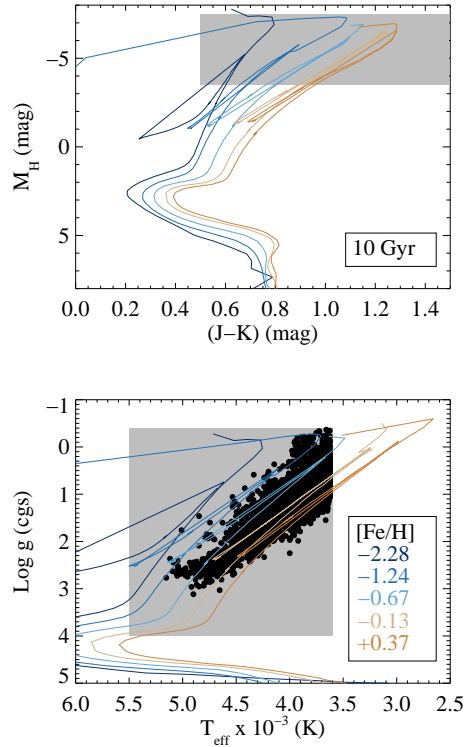}
\caption{The selected sample (black) is compared with
theoretical isochrones (Marigo et al. 2008) for an age of 10\,Gyr and five different metallicities. The gray regions indicate 
the selected parameter space, defined by brightness (for a typical bulge distance of 8.0\,kpc from the Sun) and color limits, 
as well as calibration restrictions. }
\end{center}
\label{isochr}
\end{figure}

Large stellar samples of giants spanning different regions of the bulge are ideal 
for exploring this Galactic component. That was the main motivation for including the observations here
described in the APOGEE survey. However, such samples can suffer from selection biases associated with 
target selection and/or limitations in the spectroscopic analysis, which may skew the derived parameters and 
overall sample statistics. For a typical bulge  distance and age (8\,kpc and 10\,Gyr), the APOGEE data base 
samples only the top of the red and asymptotic giant branches (RGB and AGB, see Figure\,\ref{isochr}).  
In the present work, the culled stars comprise $\sim 63\%$ of the 20,707 survey giants 
($\log{g} \leqslant 3.8$) in the 83 fields considered. The most common rejection factor after the distance cut, was having poor {\teff} 
estimates, mainly due to the proximity to the cool edge of our model grids ({\teff}$=3500$ K), 
which affects 17\% of the 20,707 stars. Among these cool stars lost from the sample, we are preferentially
missing the most metal-rich stars, which could distort somewhat the high-end of our inferred metallicity distributions.

We  make a quantitative estimation of the biases present in the APOGEE bulge sample by using  
the \citet{Chabrier01} IMF and integrating it for different \citet{Marigo08} isochrones to identify which fraction of 
any given mono-age and mono-metallicity population in the bulge makes the APOGEE cut: 
$7 \leqslant H \leqslant 11$, $(J-K_{\mathrm{s}})\geqslant 0.5$, $3600 \leqslant$ {\teff} $\ \leqslant 5500$\,K, 
and $-0.4\leqslant $ {\logg} $\ \leqslant 4.0$.  The integral is computed for multiple isochrones with a 
relevant range in age (5 and 10 Gyr), and metallicity (approximately between $-2.0$ and $+0.5$). Variations 
in extinction ($A_K=0$ and 1) and distance (5 and 8 kpc) are also considered. The fraction of stars observed for each
case is computed as the fraction between the integral over the part of the isochrone that satisfies the APOGEE cut ($\int_\mathrm{c}{\xi}dM$)
and the same integral over the entire range of stellar mas for RGB and AGB stars ($\int_\mathrm{g}{\xi}dM$).

\begin{figure}
\figurenum{3}
\begin{center}
\includegraphics[angle=0,trim=1.0cm 1cm 3cm 2cm,clip,scale=0.36]{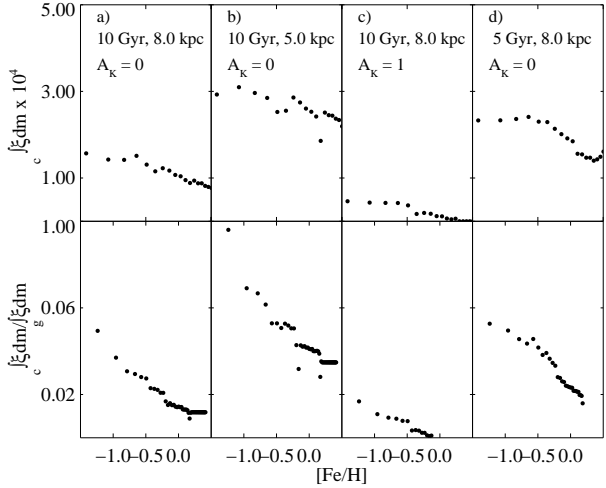}
\caption{Integrated IMF over the isochrones (top) in the APOGEE brightness, color, \teff\ and \logg\ ranges, and its fraction (bottom) in the entire RGB-AGB range. Values are shown as a function of metallicity for different combinations of distance, extinction, and age. The results of the integrations are normalized to a stellar population of one solar mass.} 
\end{center}
\label{fractions}
\end{figure}

A few examples of our APOGEE targeting efficiency estimates are shown in Figure\,\ref{fractions}.  
The top row of panels shows the integral of the IMF over the window defined by the APOGEE cut. 
The bottom row of panels in the figure shows the relative fraction of stars that make the cut. 
A more complete coverage of the entire RGB-AGB is achieved at shorter distances, and the same
is true for lower values of the interstellar extinction.

As discussed above, metal-poor stars are expected to have a better sampling than 
metal-rich stars because of the {\teff} cut:  at the highest metallicities (\feh$\geqslant+0.2$), 
the brightest parts of the RGB and AGB become cooler than the specified \teff\ limit.
For a 10 Gyr-old bulge population the APOGEE sample would consist of 0.9--1\,M$_\sun$ stars in the RGB and AGB phases. For a  5\,Gyr-old bulge population, the APOGEE observations would  include instead 1.0--1.3\,M$_\sun$ stars. The APOGEE bulge sample cuts favor low (\feh$\sim-1.3$) over high metallicities (\feh$>0$). 

Stars in the red giant branch are statistically better represented at younger ages, 
closer distances, and/or lower extinctions. Overall, the highest metallicities will underrepresented by some
fraction around 30 \% relative to the stars at [Fe/H] $\sim -1$ (up to 90\% for the most
distant regions with high extinction). Our APOGEE-based RGB-AGB metallicity distribution 
functions (MDFs, Section\,\ref{MDFs}) will be affected by these issues, and the derived metallicity 
distributions will be distorted, but since departures from the truth distributions should be similar across regions at similar distances, {\it relative variations} across the bulge are much more robust, 
and we focus on those in the present analysis.

\section{Distances}
\label{samdist}

\begin{figure}
\figurenum{4}
\begin{center}
\includegraphics[trim=5cm 5.4cm 3.0cm 0cm,angle=0,scale=0.18,clip]{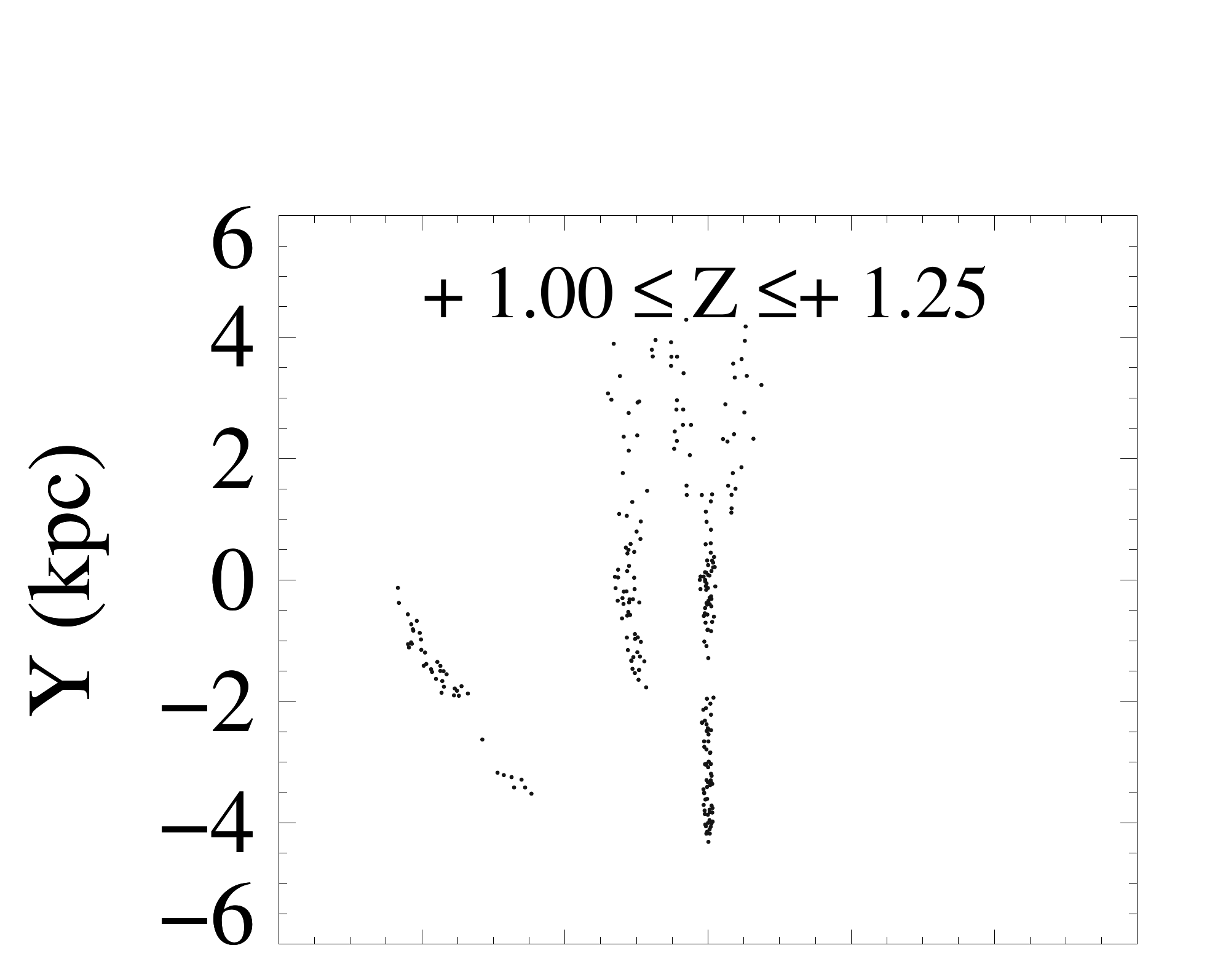}
\includegraphics[trim=9.3cm 5.4cm 3.0cm 0cm,angle=0,scale=0.18,clip]{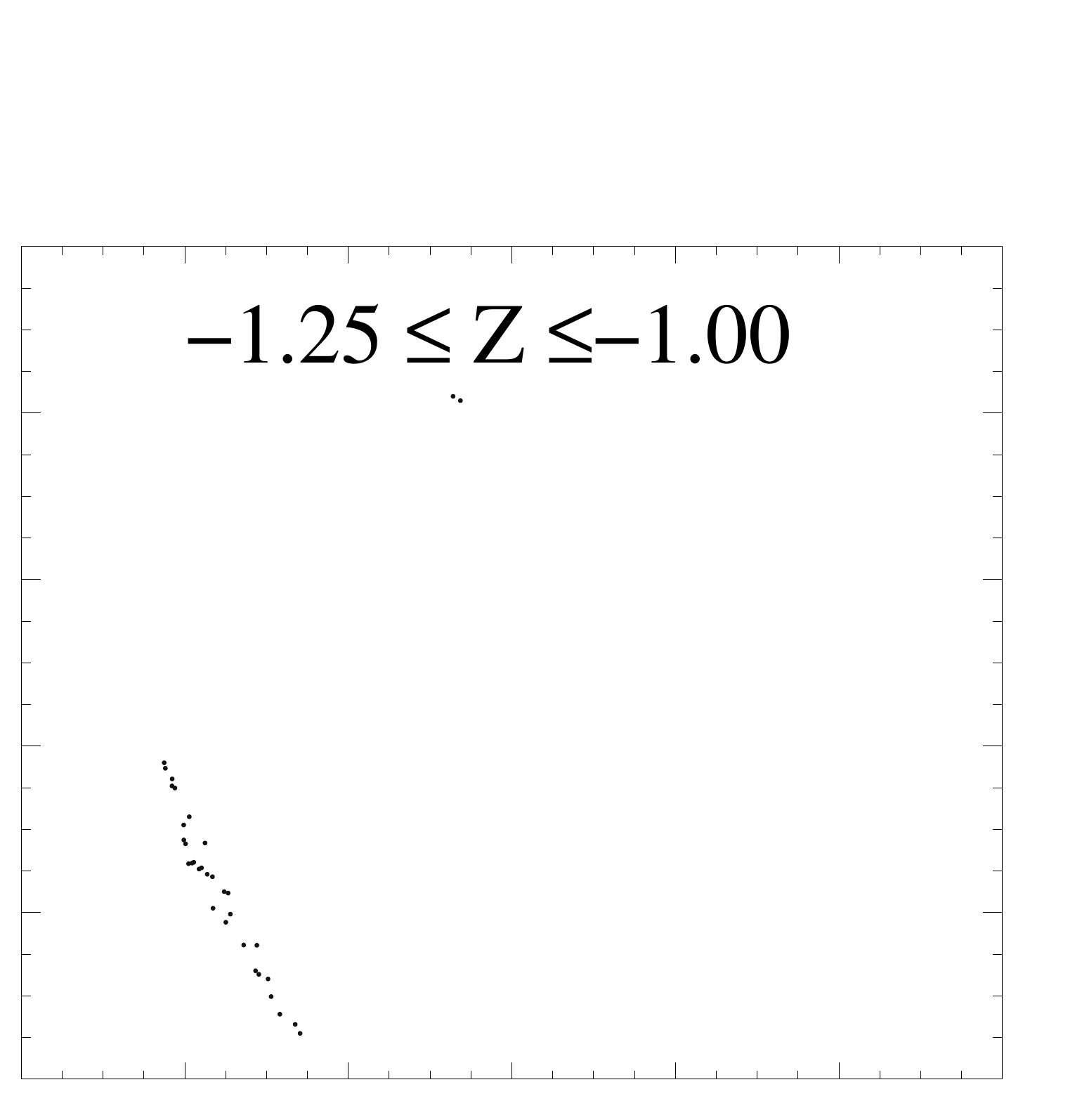}\\
\includegraphics[trim=5cm 5.4cm 3cm 3cm,angle=0,scale=0.18,clip]{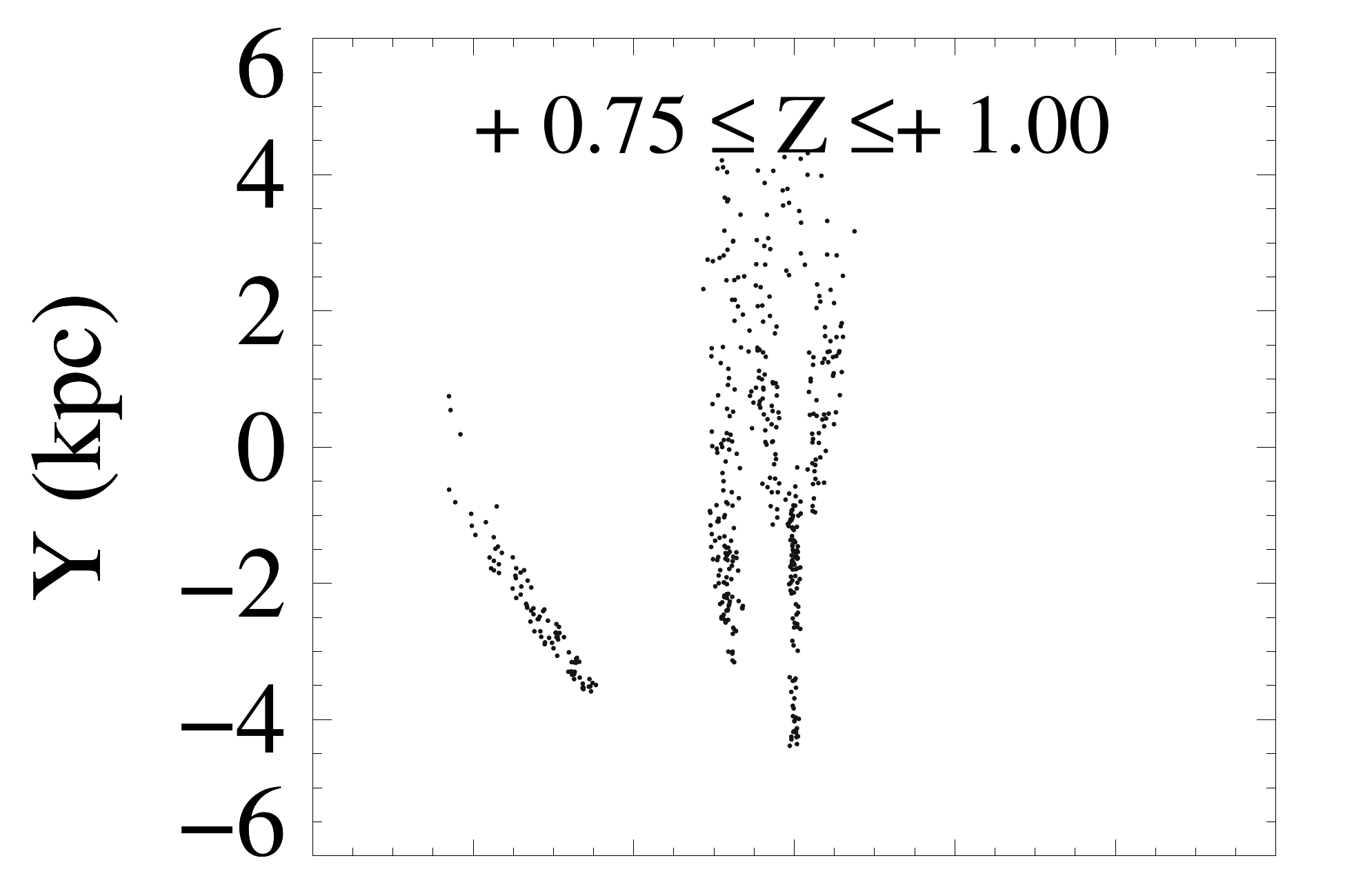}
\includegraphics[trim=9.3cm 5.4cm 3cm 3cm,angle=0,scale=0.18,clip]{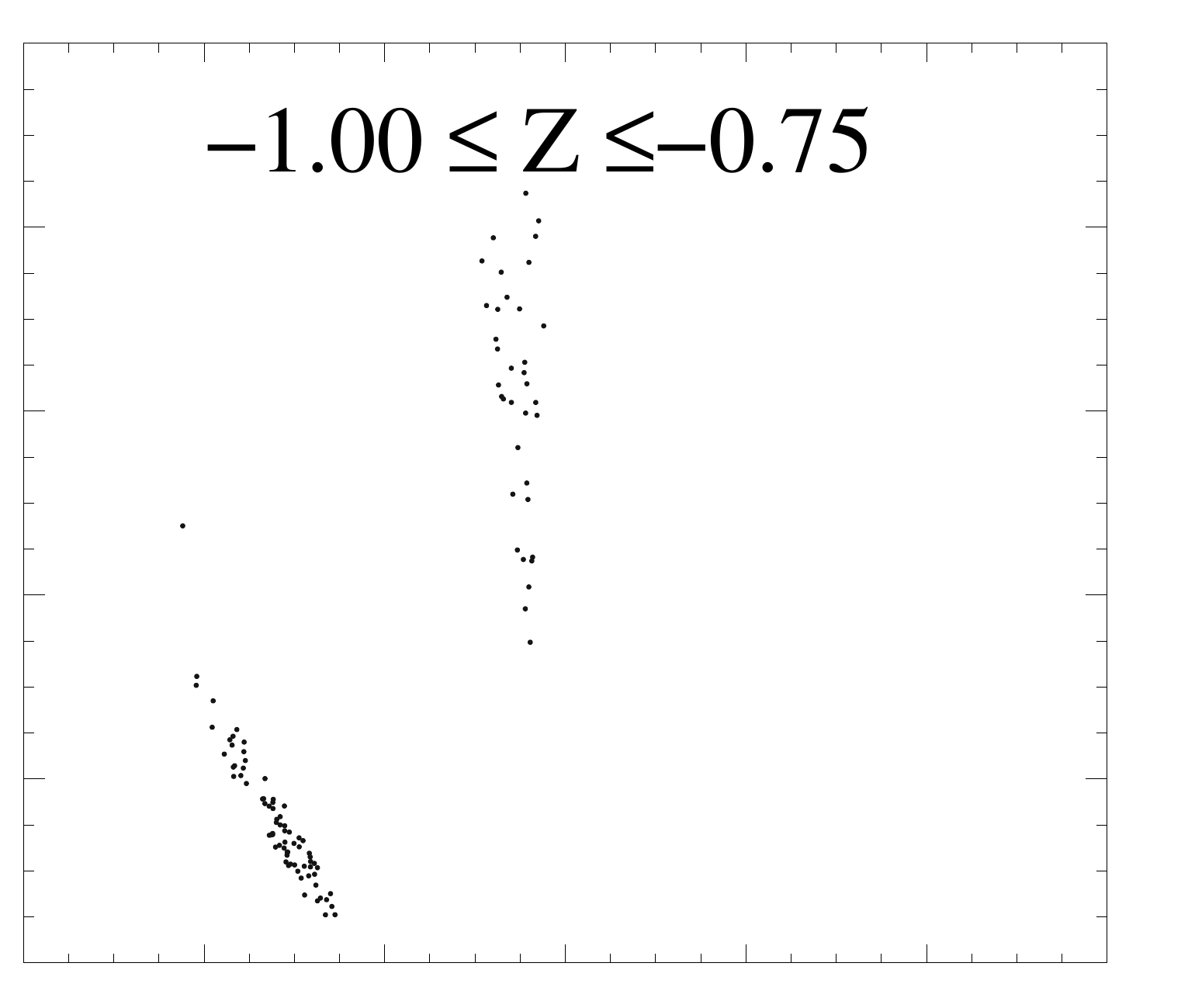}\\
\includegraphics[trim=5cm 5.4cm 3cm 3cm,angle=0,scale=0.18,clip]{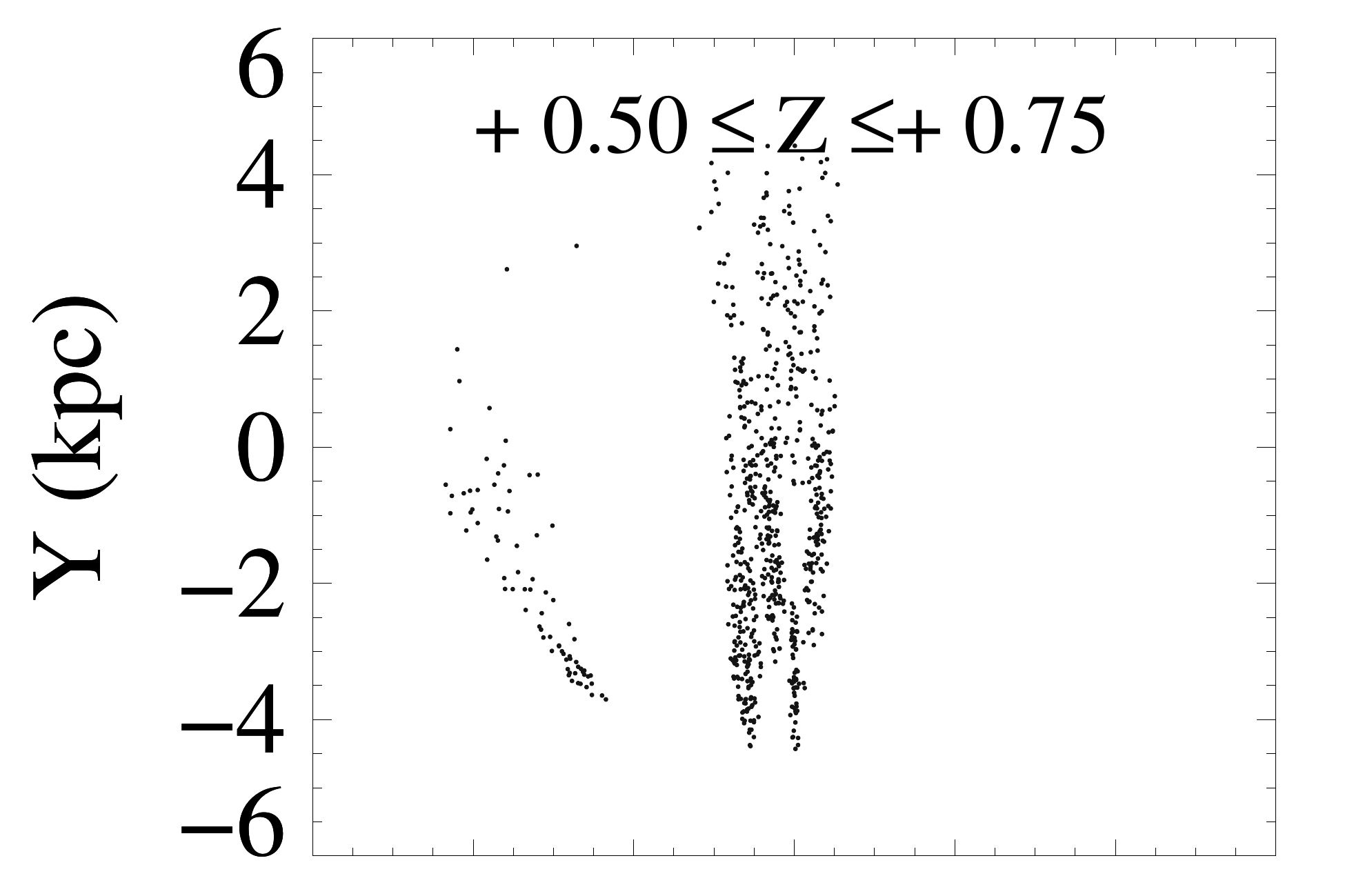}
\includegraphics[trim=9.3cm 5.4cm 3cm 3cm,angle=0,scale=0.18,clip]{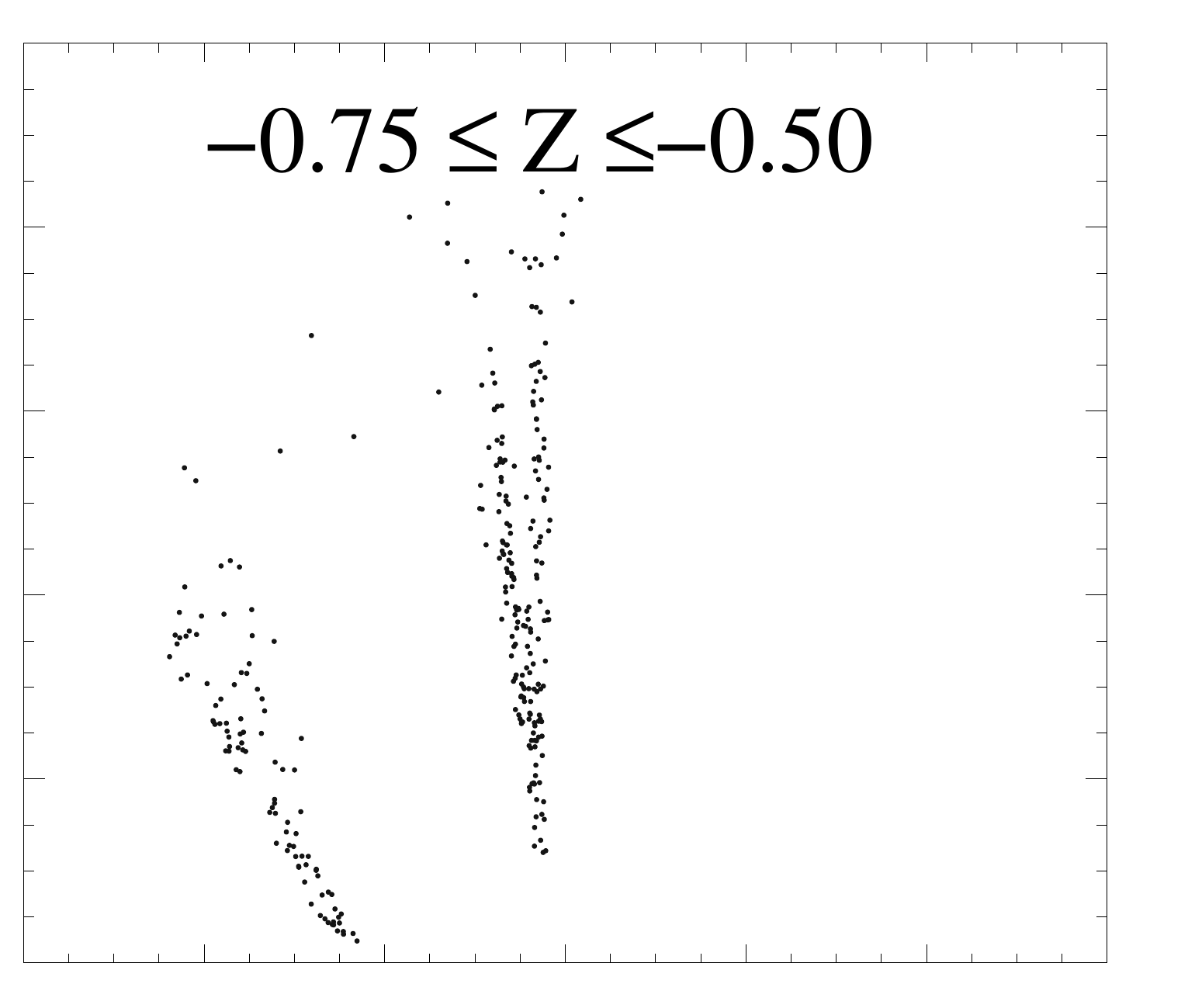}\\
\includegraphics[trim=5cm 5.4cm 3cm 3cm,angle=0,scale=0.18,clip]{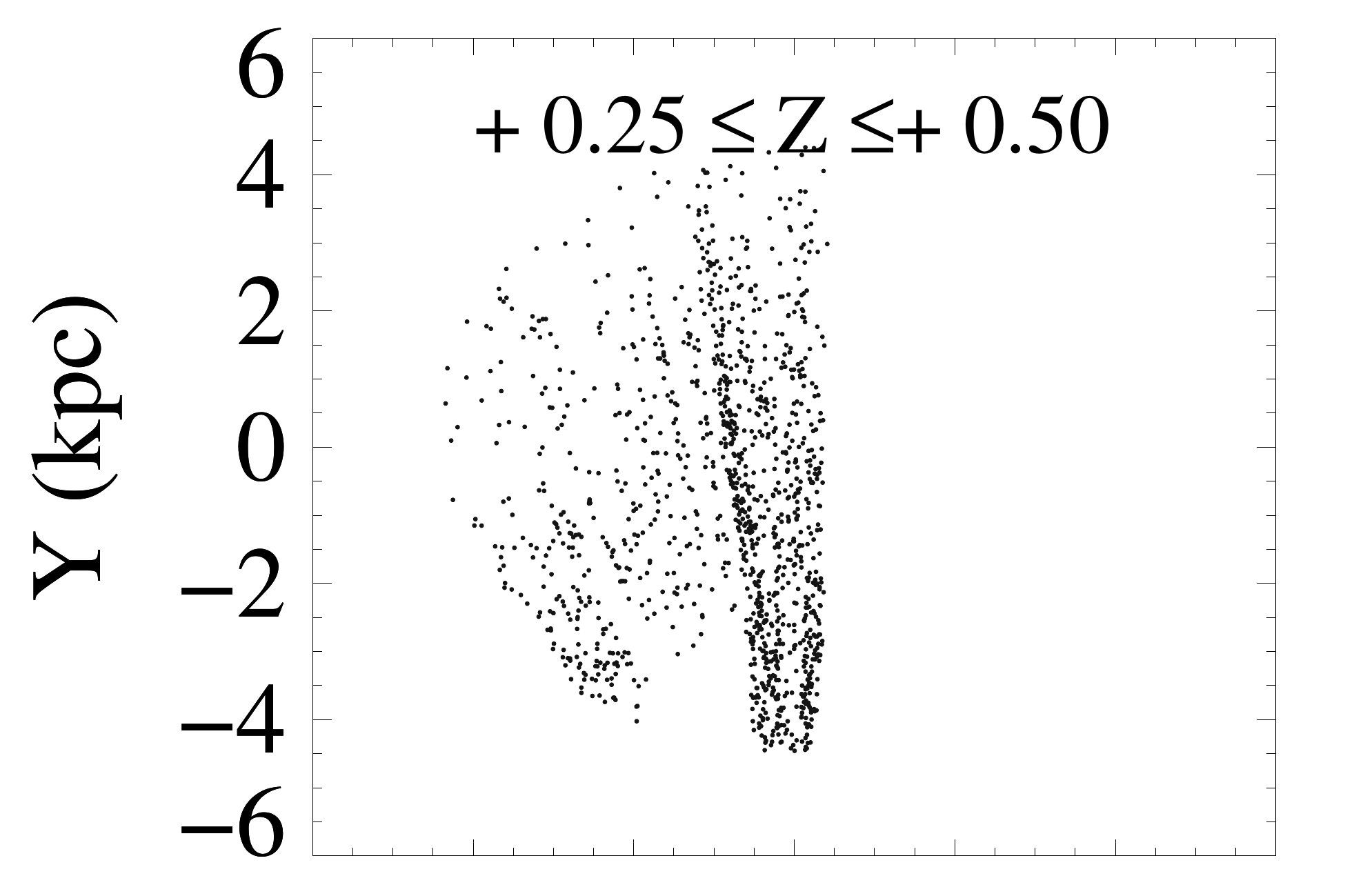}
\includegraphics[trim=9.3cm 5.4cm 3cm 3cm,angle=0,scale=0.18,clip]{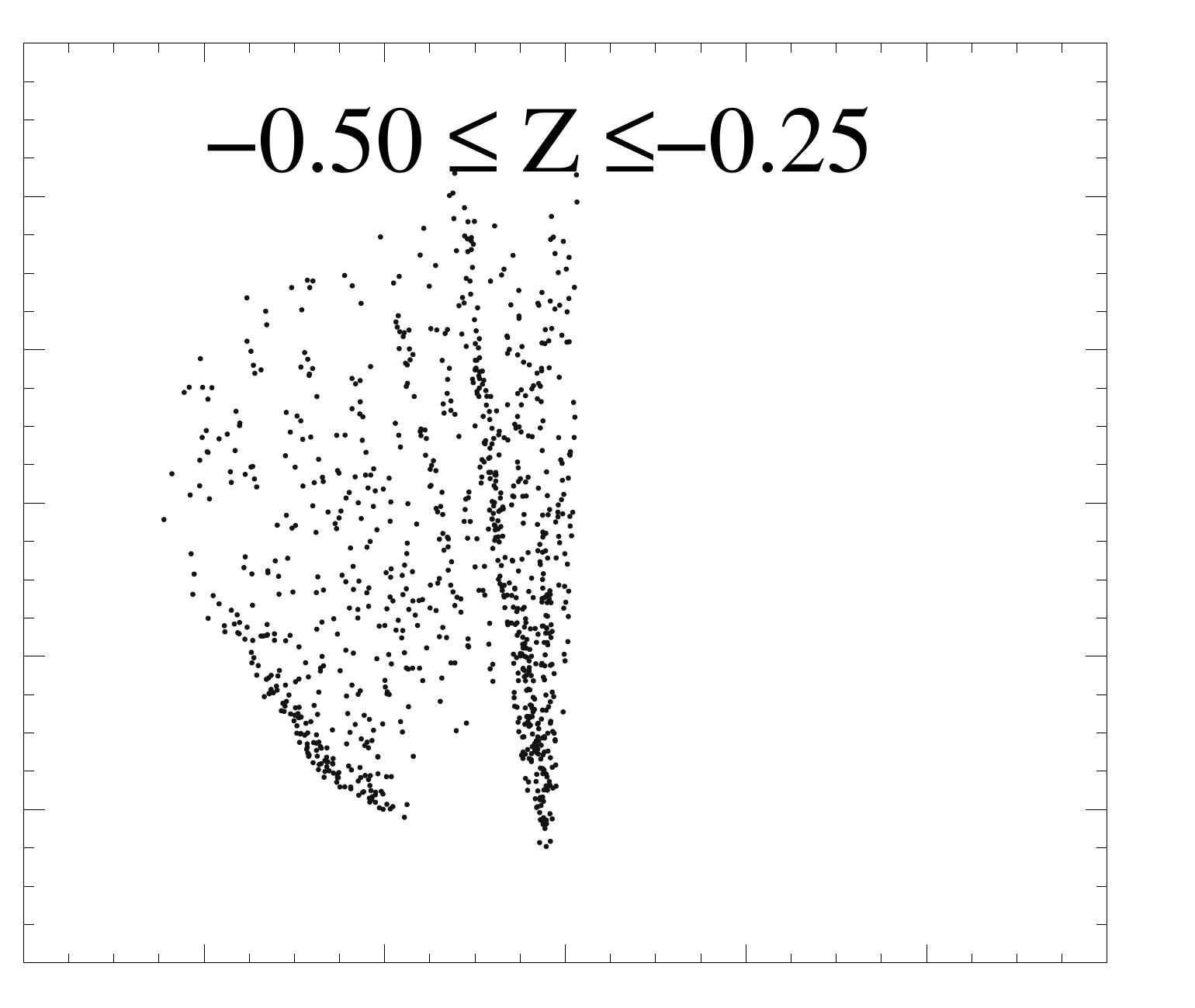}\\
\includegraphics[trim=5cm 2cm 3cm 3cm,angle=0,scale=0.18,clip]{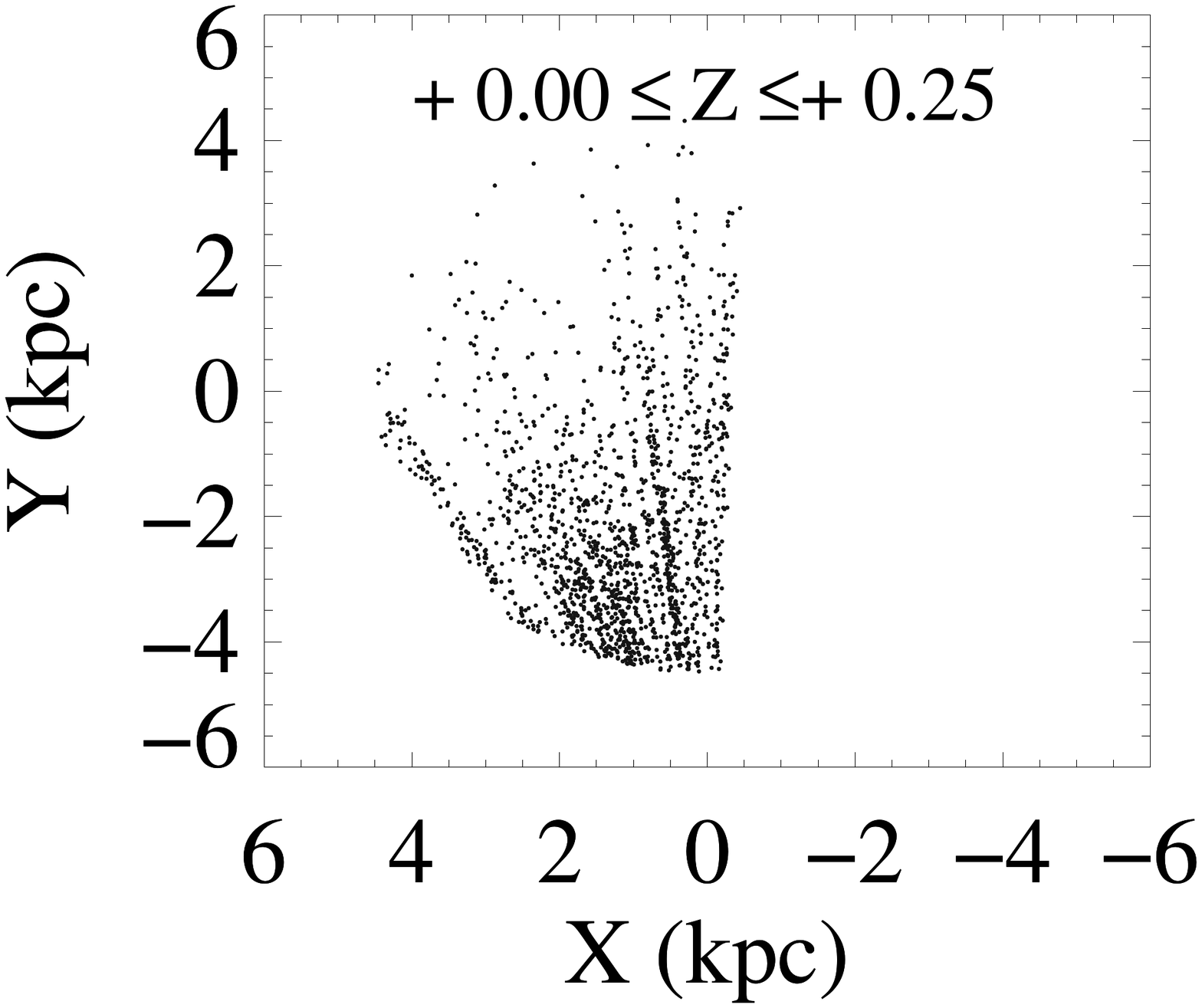}
\includegraphics[trim=9.3cm 2cm 3.0cm 3cm,angle=0,scale=0.18,clip]{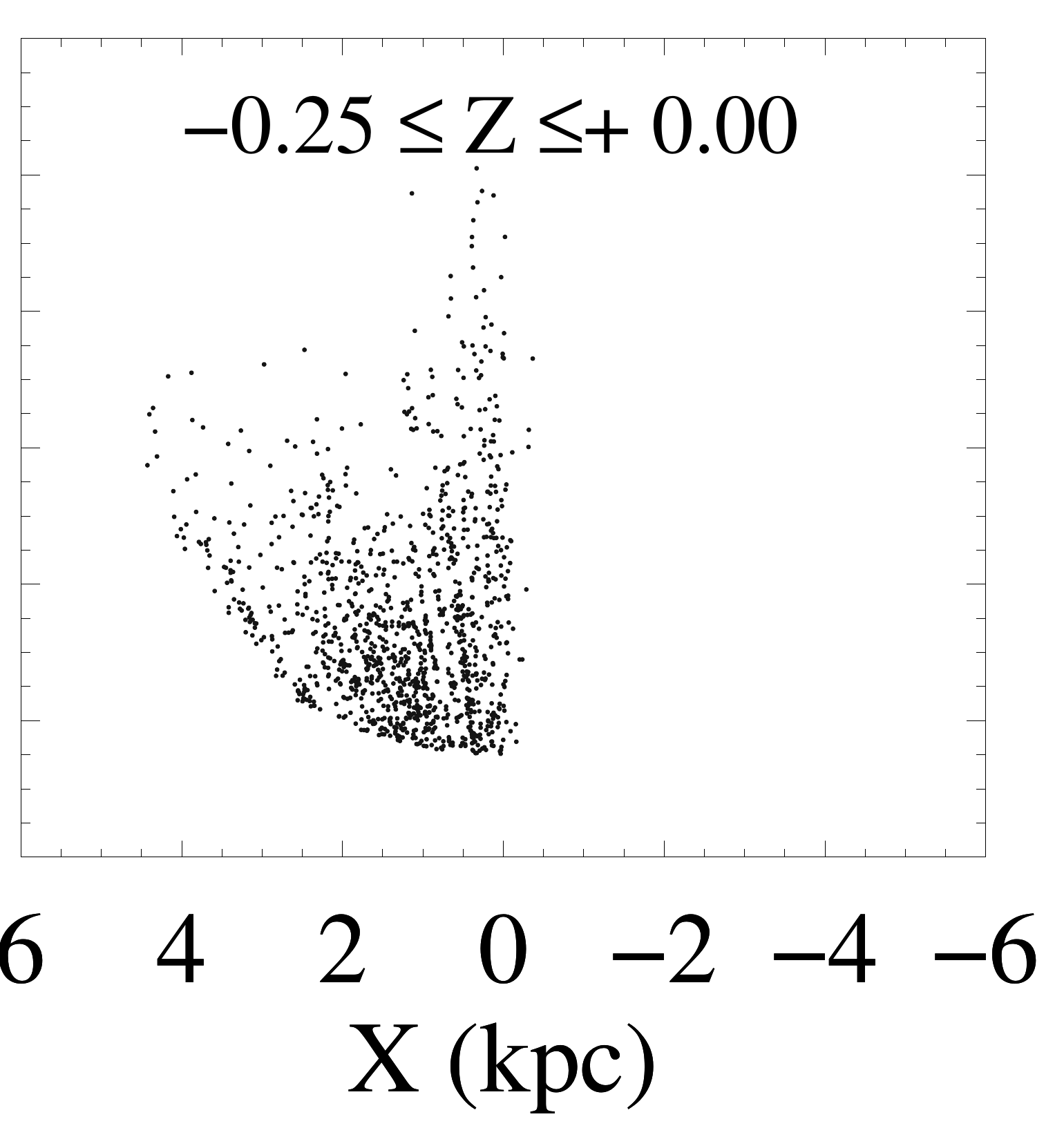}\\
\caption{Spatial distribution of the stellar sample in Galactic Cartesian coordinates $XYZ$.}
\end{center}
\label{dmaps}
\end{figure}
For the definition of the bulge sample, we adopted a solar Galactocentric distance of 8\,kpc, 
and the distance estimates from \cite{Hayden15} 
with a limit of 4.5\,kpc in Galactocentric distance ($d_{GC}$). This limit is intended to restrict the 
sample to the inner Galaxy. \cite{Hayden15} used a Bayesian method that assumed three stellar density priors (bulge, disk, halo) 
and relied upon reddening estimates \citep{Zasowski13} as well as Padova isochrones \citep{Bressan12} 
to generate stellar distances of $\sim20\%$ accuracy.  Specifically, Hayden et al. computed 
Probability Density Functions (PDFs) for various combinations of metallicity, mass, and age based 
on the probability of belonging to the triaxial bulge, disk, and halo, and taking into account 
the initial mass function. With these distance determinations, $\sim50\%$ of the stars observed in the 83 
fields have $d_{GC} \leqslant 4.5$\,kpc and approximately 7,000 are within 1.25\,kpc of the plane.  
Note that some foreground contamination is present in each of the fields.

\begin{figure*}
\figurenum{5}
\begin{center}
\includegraphics[angle=0,trim=2.5cm 0cm 0cm 0cm, scale=0.6,clip]{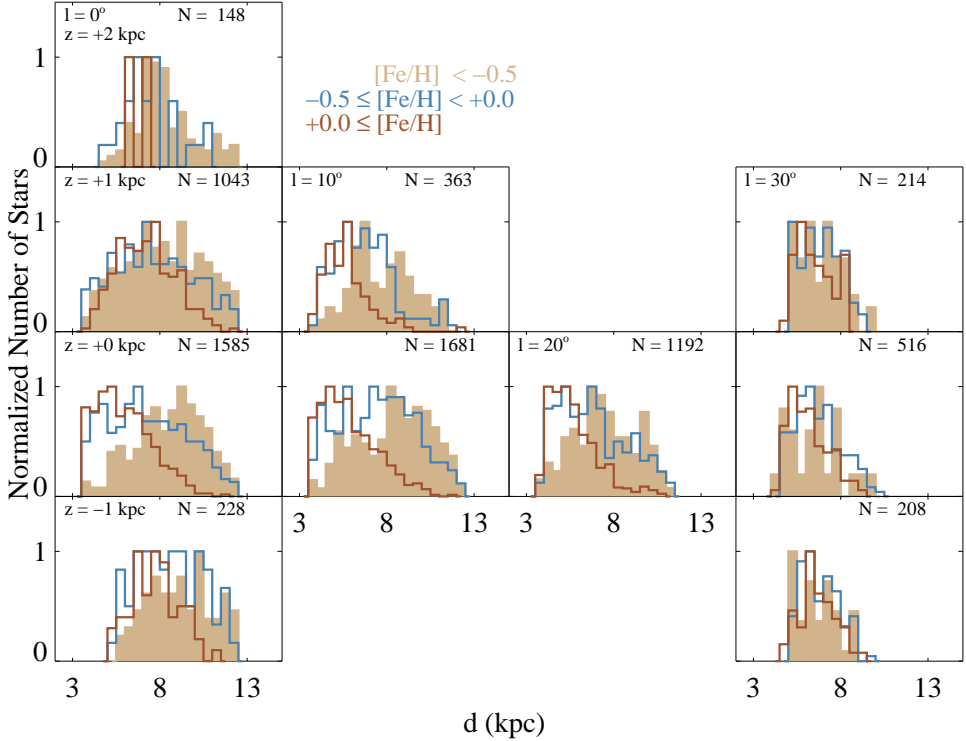}
\caption{Distance distribution (from the Sun) of the bulge sample ($d_\mathrm{GC} \leqslant 4.5$\,kpc) in bins of 0.5\,kpc 
and separated by Galactic longitude ($10\degr$ bins), heights from the midplane (1\,kpc bins), and metallicities. 
Histograms are normalized to peak values and the outline colors represent the three different metallicity groupings.}
\label{dist}
\end{center}
\end{figure*}

Figure\,\ref{dmaps} shows the spatial distribution in Galactic Cartesian coordinates of the stellar sample, while Figure\,\ref{dist} displays the distribution of their distances for  different Galactic longitudes 
and derived heights from the plane ($Z= d \sin b$).  Only regions with more than 20 stars are considered in the latter figure.
We find that APOGEE-1 samples mostly the near-side of the bulge ($d \lesssim 8$\,kpc). 
To study variations with metallicity, stars were placed into three groups: high metallicity (\feh$\ \geqslant 0.0$), 
intermediate metallicity ($-0.5\leqslant$\ \feh\ $\leqslant+0.0$), and low metallicity (\feh$\ \leqslant -0.5$).  
This grouping scheme was informed by the MDFs described in Section\,\ref{MDFs}.  
As shown in Figure\,\ref{dist}, in the midplane, the most metal-poor population tends to be 
more distant than the metal-rich population and covers a wider range of distances at low Galactic longitudes.  
The intermediate metallicity population has a median in distance distribution between these two.  
As Galactic longitude increases or proceeding higher in the bulge, the separation in distance 
between the high and low metallicity groups seems to grow smaller. 
Some of this departure at low $l$ may be attributed to the bias towards low metallicity 
stars at large distances and higher extinction, as described above.  Nevertheless, 
some of this difference may also be intrinsic to the structure of the inner Galaxy, e.g., metallicity populations of different scale-heights \citep{Robin12,Robin14}.


\section{Metallicity Maps}

\label{map}

The existing APOGEE observations provide larger bulge coverage at high spectral resolution than 
any other existing data set.
Variations in the metallicity distribution across the bulge inform about possible formation scenarios for the bulge and shed light on the link between these regions with other Galactic 
components/stellar populations (e.g., disk, bar, and halo). 
We have used the APOGEE metallicities to calculate the median value at various positions along the inner Galaxy.
This is done after binning the sample in Galactocentric ($X$, $Y$, $Z$) with sizes of 0.5\,kpc, 0.5\,kpc, and 0.25\,kpc, respectively. 
Previous results for the bulge have suggested the presence of multiple metallicity components of 
different relative contributions \citep[e.g.,][]{Ness13}. Therefore, characterizing these distributions by medians rather than means 
was adopted because the former have lower sensitivity to outliers and thus gives a more robust representation 
of the contributing metallicities. 

\begin{figure}
\figurenum{6}
\begin{center}
\includegraphics[trim=0cm 4.2cm 7.5cm 0cm,angle=0,scale=0.22,clip]{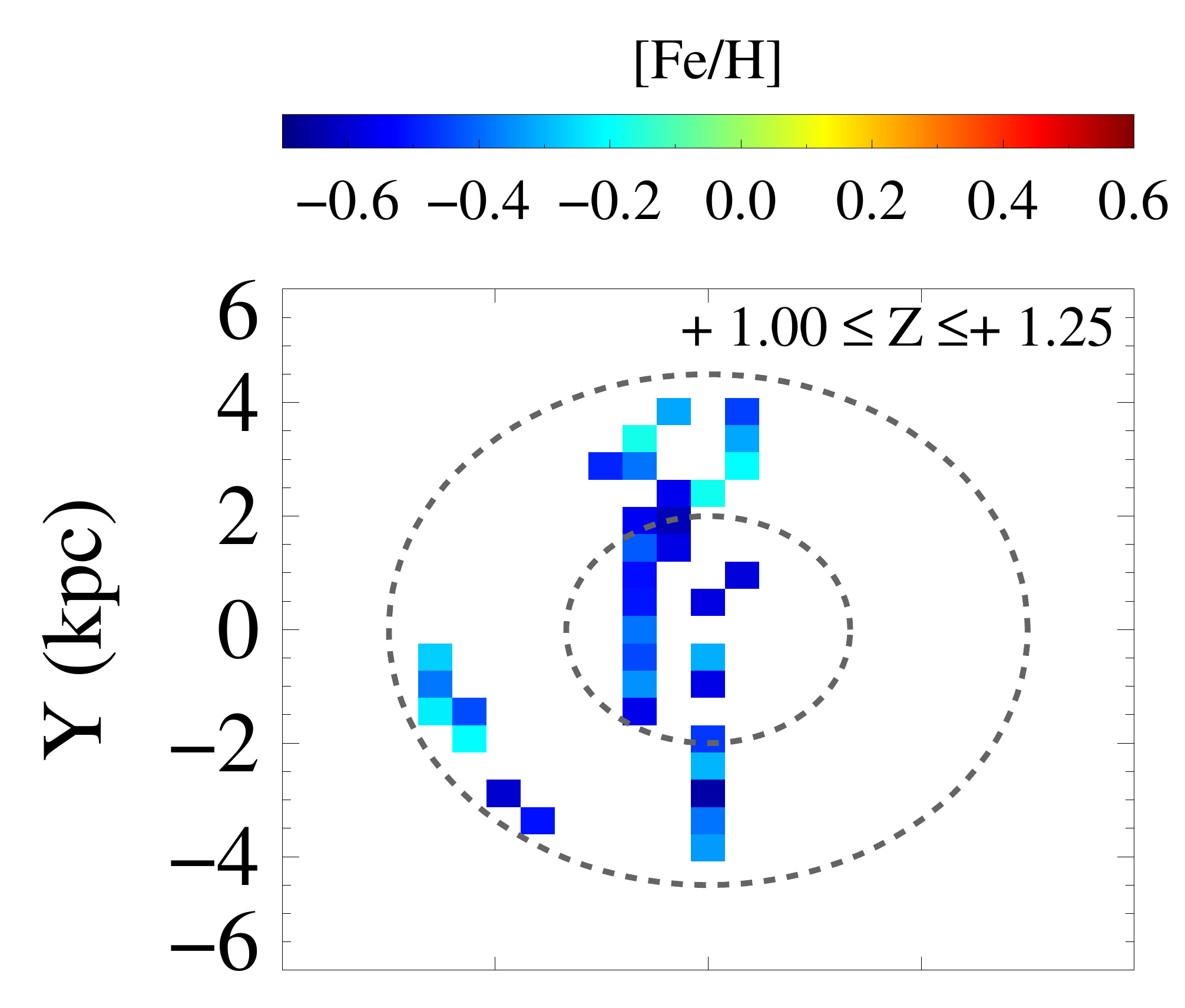}
\includegraphics[trim=4.6cm 4.2cm 7.5cm 0cm,angle=0,scale=0.22,clip]{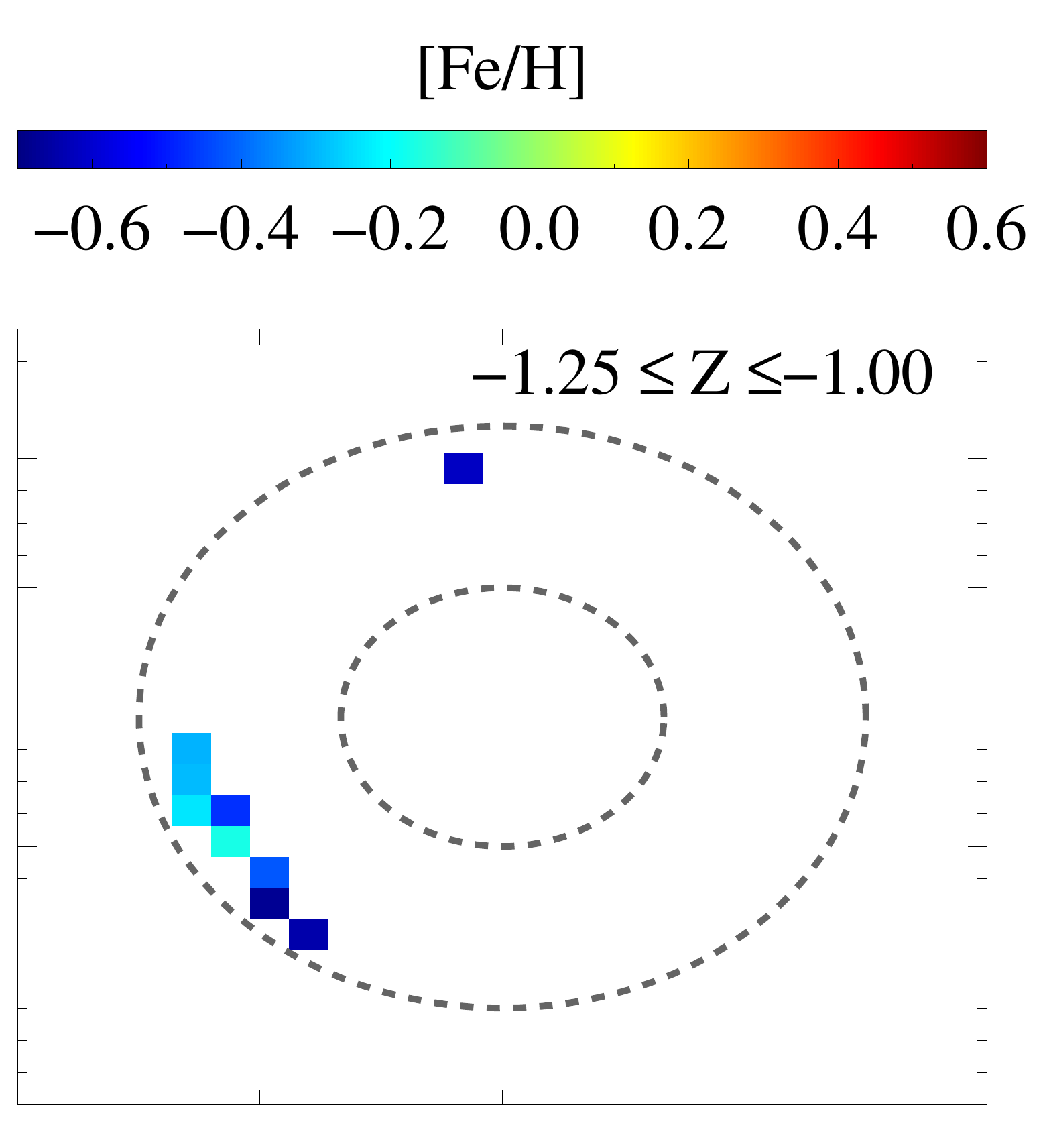}\\
\includegraphics[trim=0cm 4.2cm 7.5cm 5.0cm,angle=0,scale=0.22,clip]{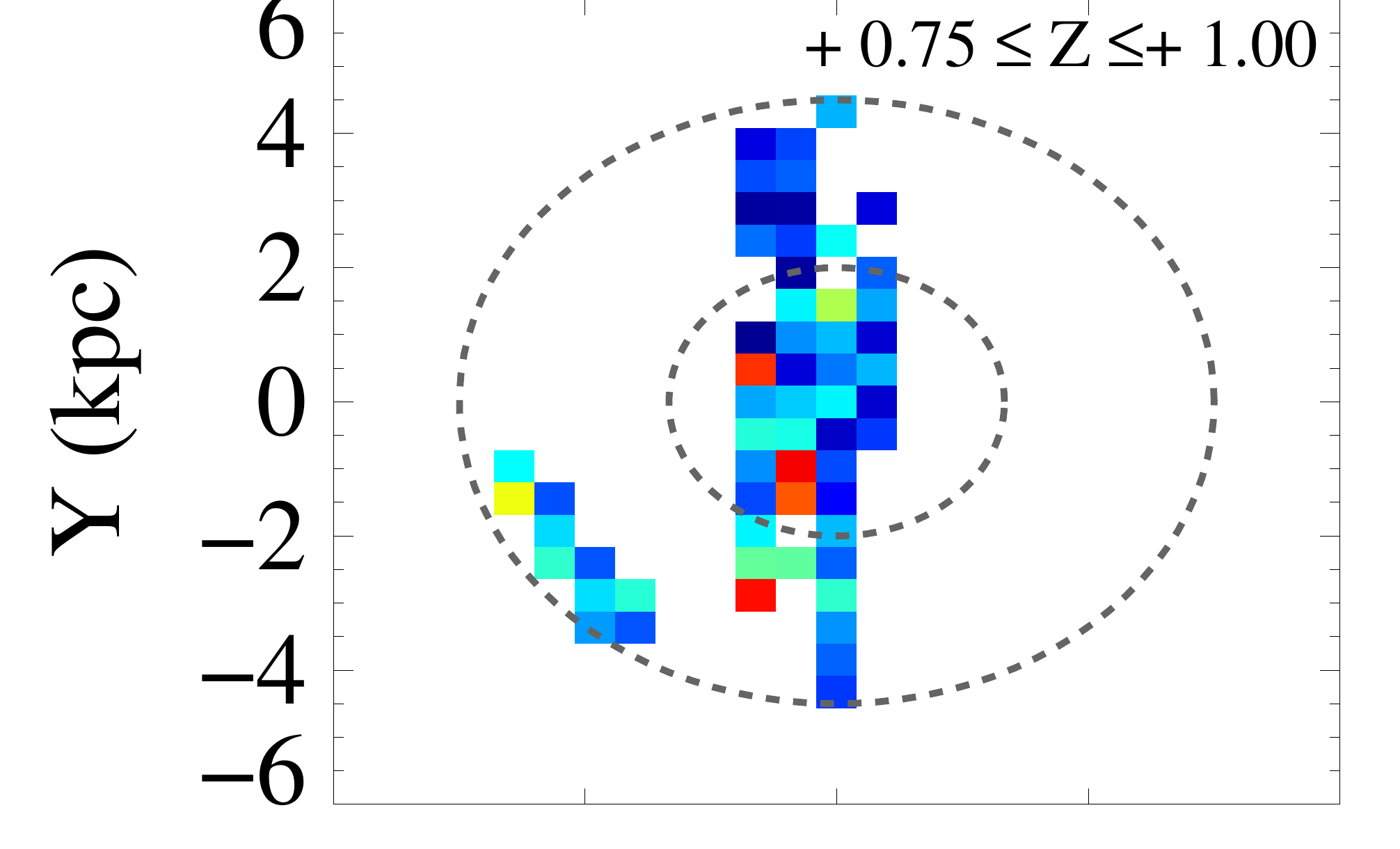}
\includegraphics[trim=4.6cm 4.2cm 7.5cm 5.0cm,angle=0,scale=0.22,clip]{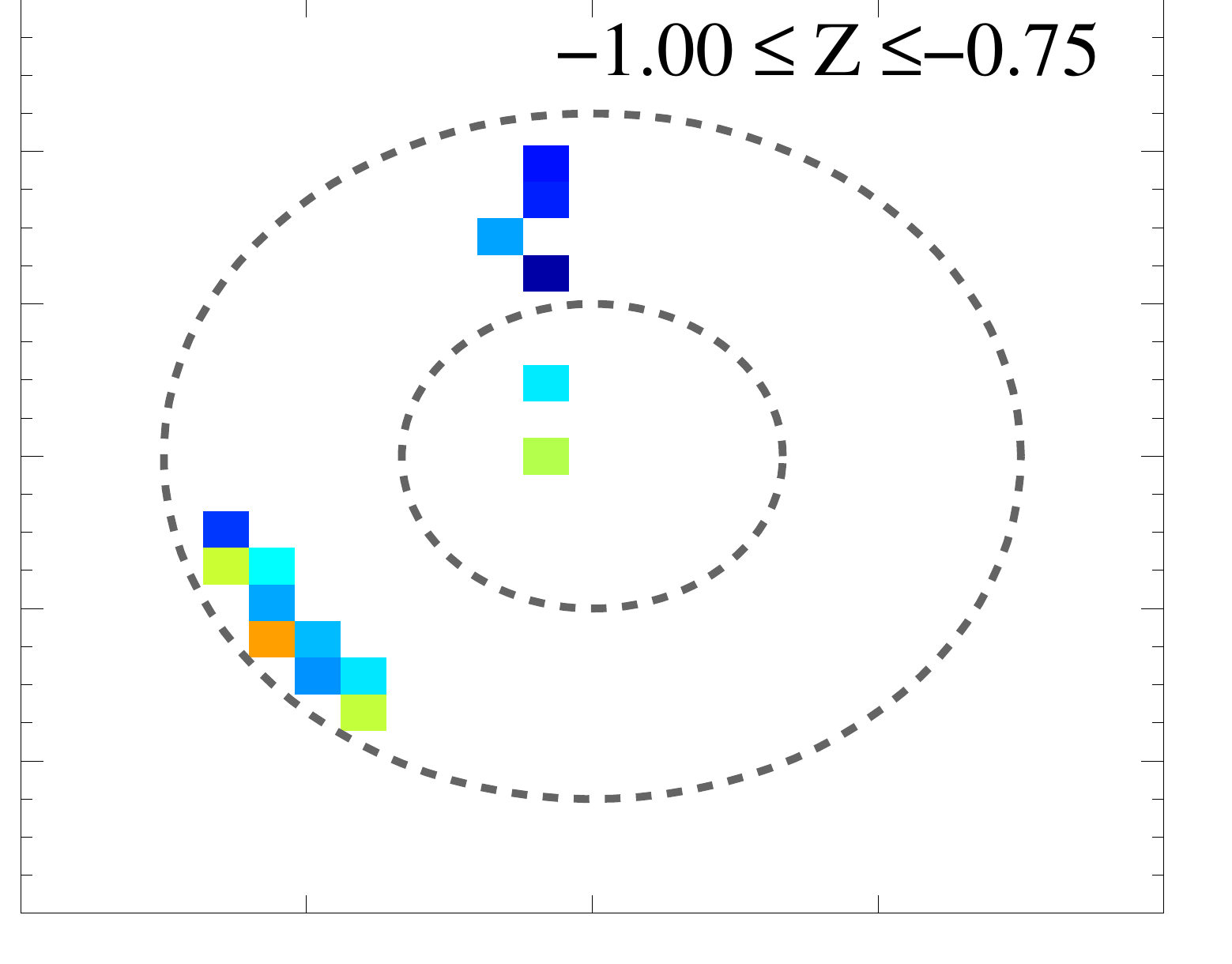}\\
\includegraphics[trim=0cm 4.2cm 7.5cm 5.0cm,angle=0,scale=0.22,clip]{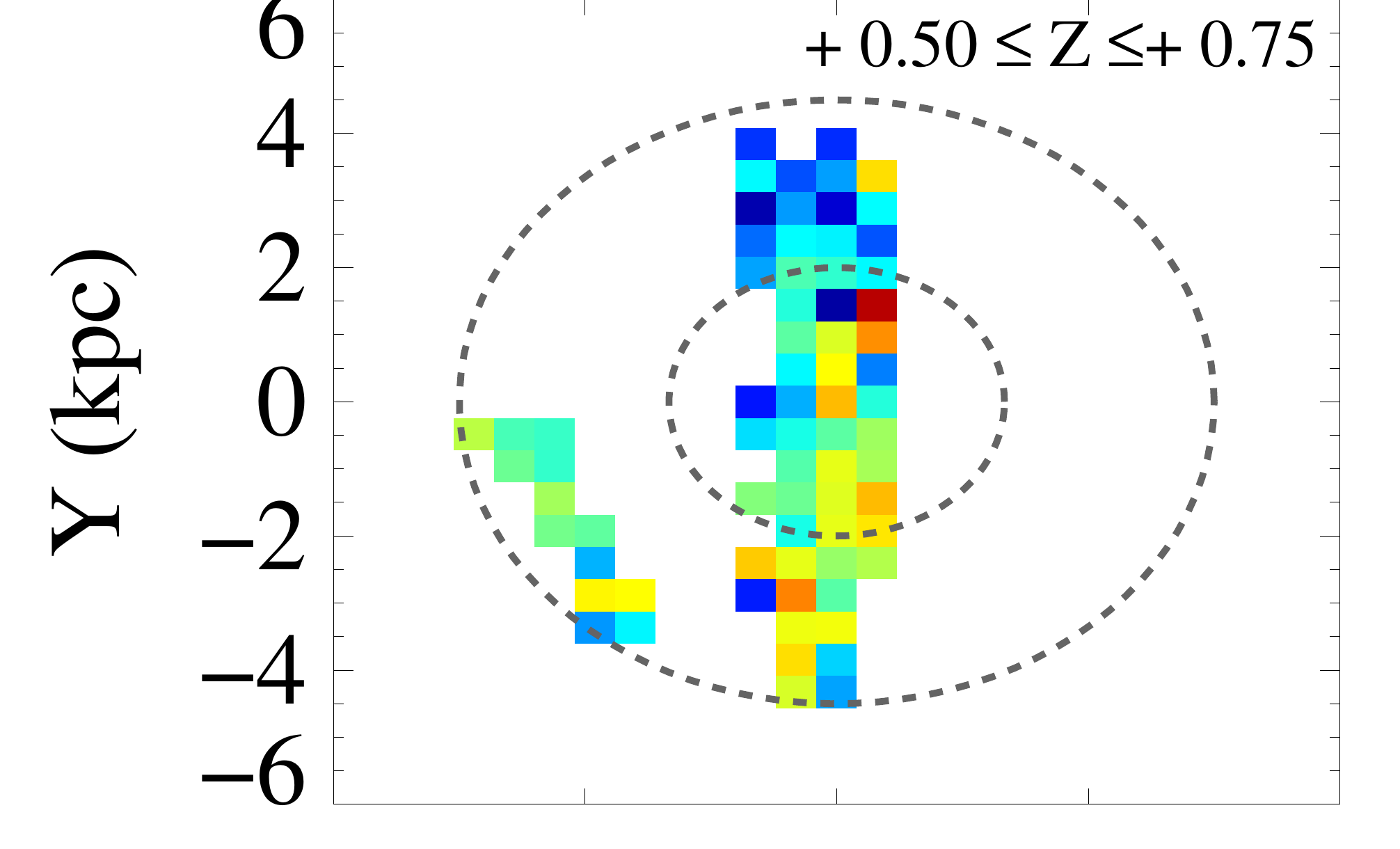}
\includegraphics[trim=4.6cm 4.2cm 7.5cm 5.0cm,angle=0,scale=0.22,clip]{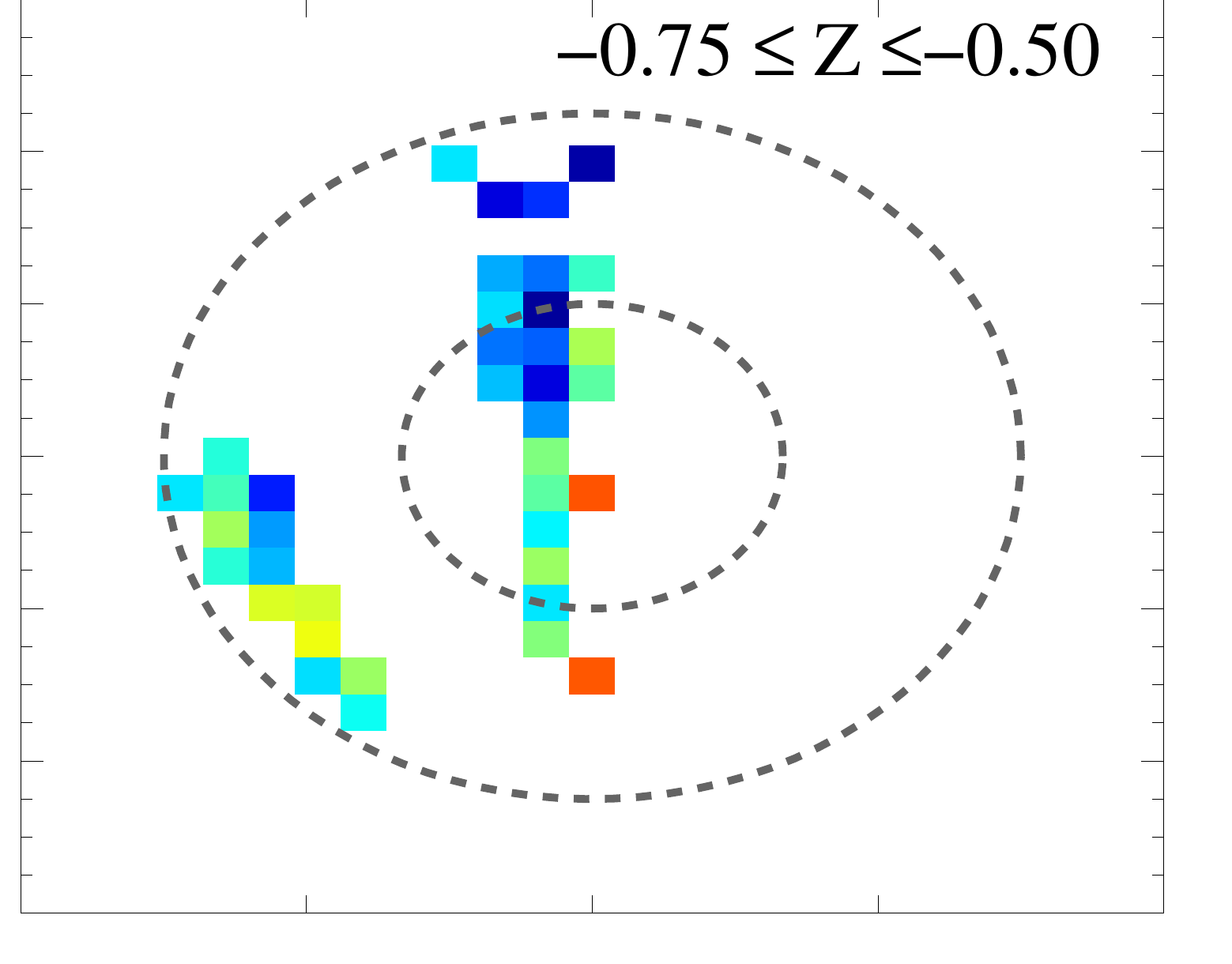}\\
\includegraphics[trim=0cm 4.2cm 7.5cm 5.0cm,angle=0,scale=0.22,clip]{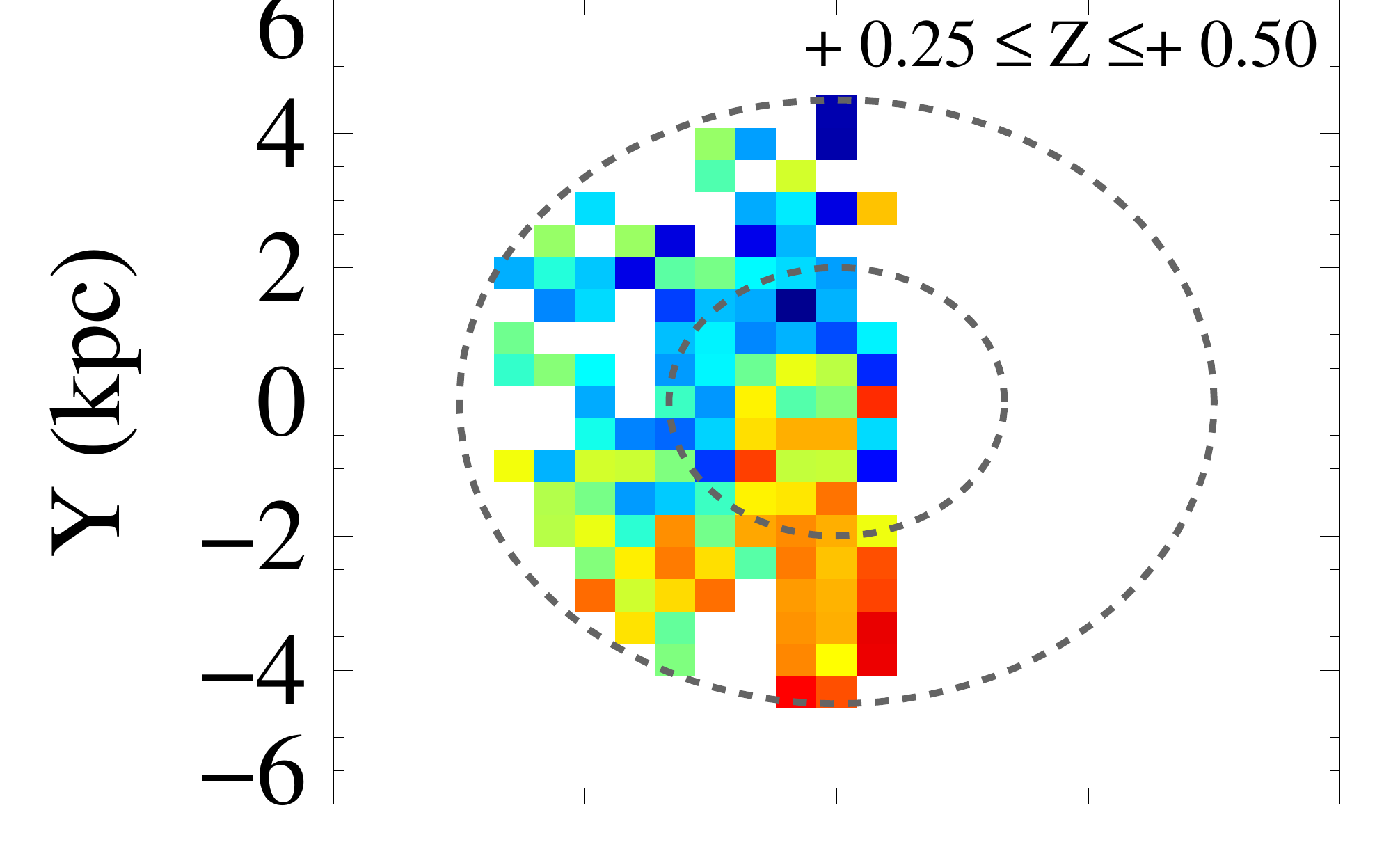}
\includegraphics[trim=4.6cm 4.2cm 7.5cm 5.0cm,angle=0,scale=0.22,clip]{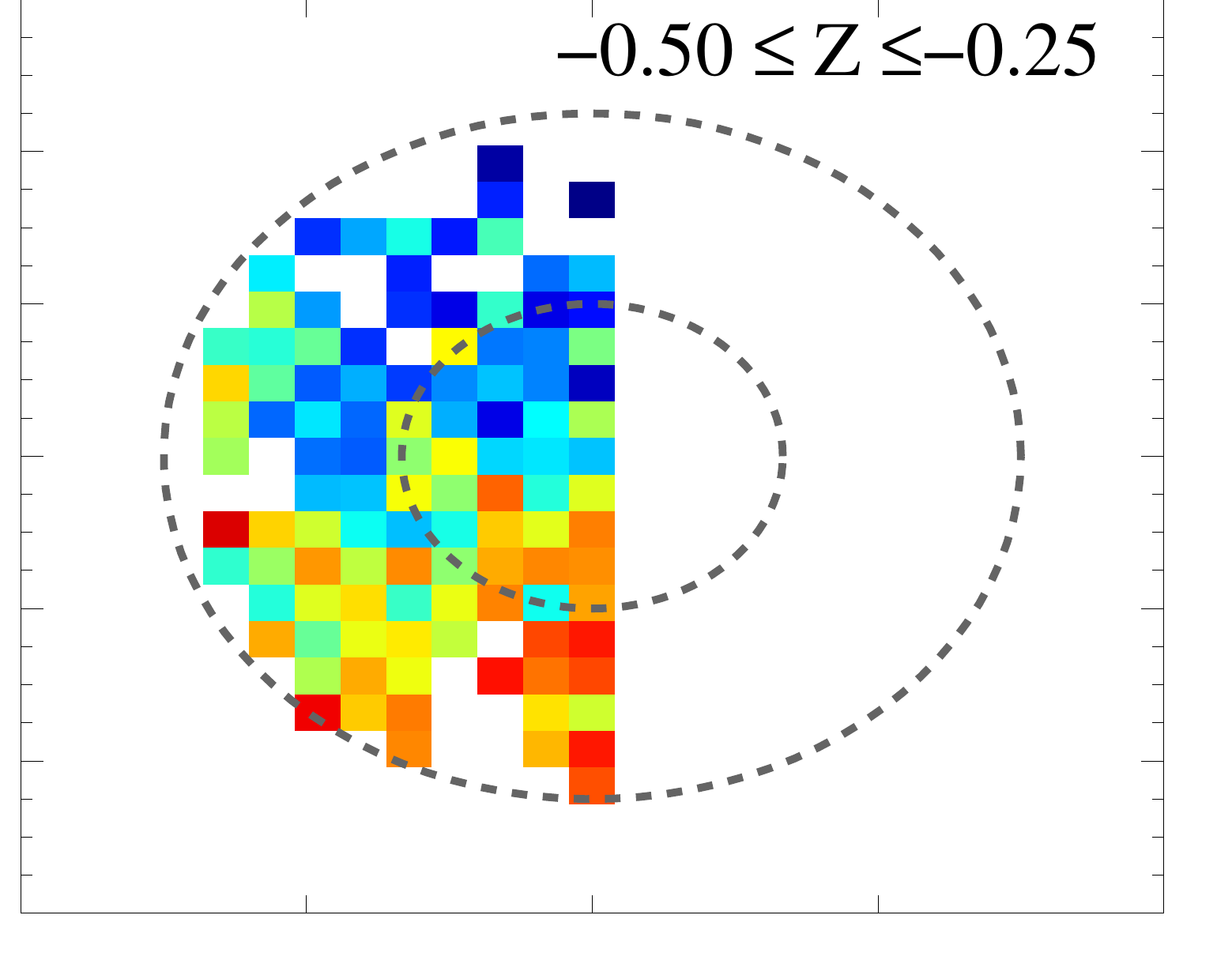}\\
\includegraphics[trim=0cm 1.cm 7.5cm 5.0cm,angle=0,scale=0.22,clip]{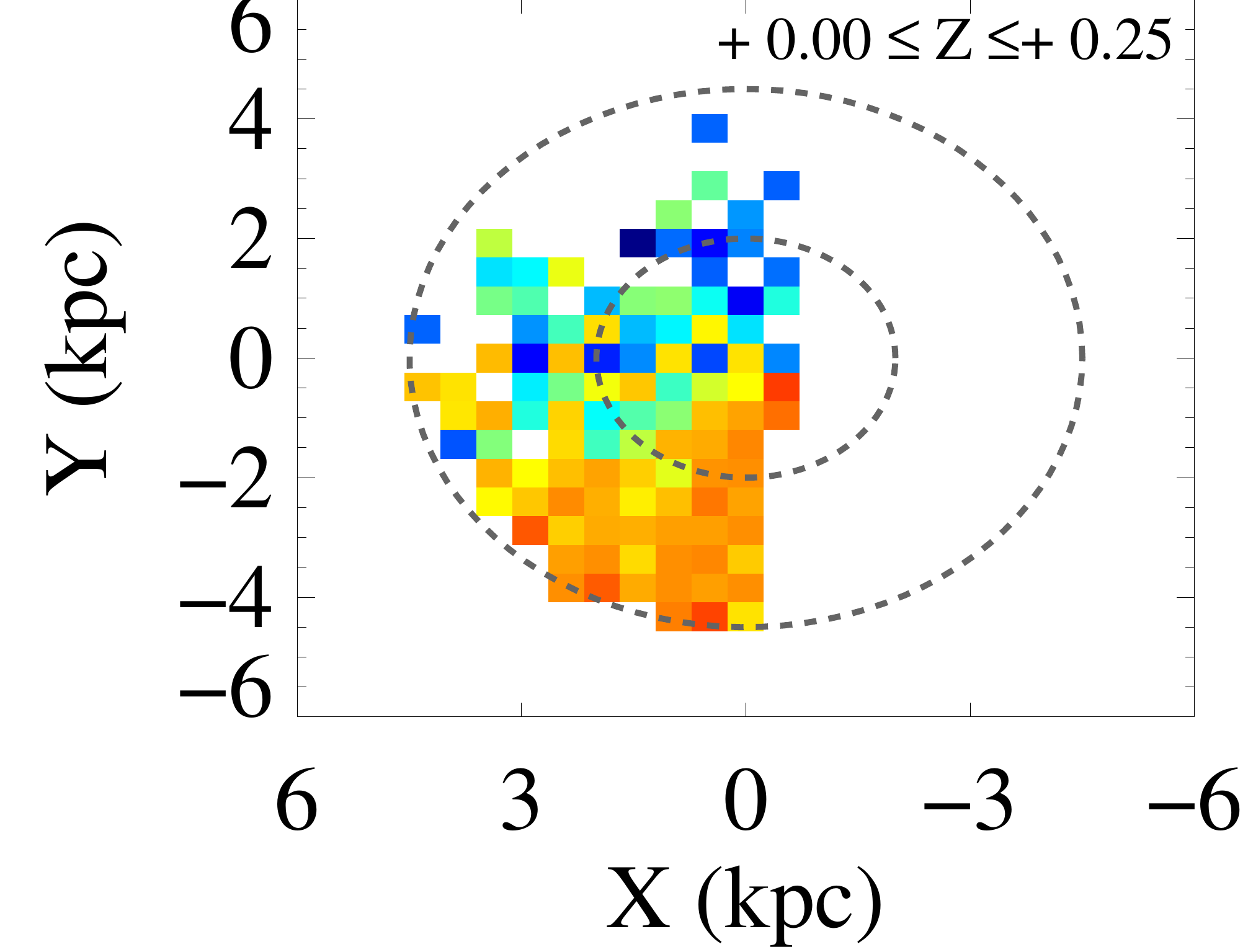}
\includegraphics[trim=4.6cm 1.cm 7.5cm 5.0cm,angle=0,scale=0.22,clip]{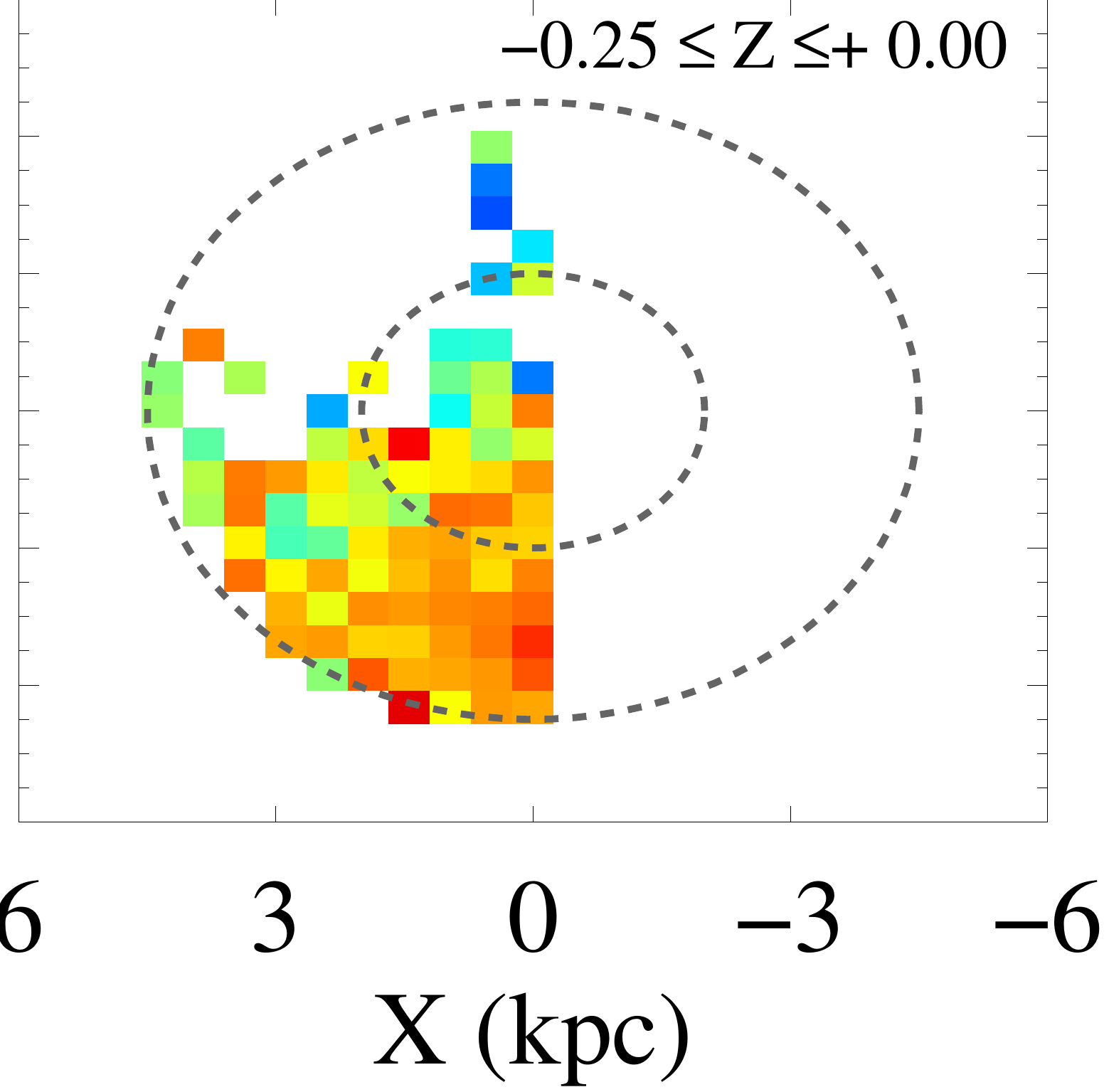}\\
\caption{Metallicity maps for the APOGEE bulge sample at different heights from the midplane and in $X$- and $Y$-bins of 0.5\,kpc. The Sun is at ($X$, $Y)=(0$, $-8)$. The color denotes metallicity and the dotted circles represent $R_{\rm{GC}}$ of 2\,kpc and 4.5\,kpc. A minimum metallicity of $-0.7$ is assigned for distinction.}
\end{center}
\label{mmaps}
\end{figure}

Figure\,\ref{mmaps} shows ($X$, $Y$) maps of the different median metallicity as a function of 
height ($Z$) from the Galactic plane.
Typical errors for the median metallicities were found to be $\sim0.05-0.10$\,dex and were estimated 
from bootstrapping simulations. In Figure\,\ref{mmaps}, the number of stars does vary with position (from a typical value of five beyond the GC or far from the midplane, to a few tens at our-side of the bulge).


 \subsection{A Metal-Rich Bulge at Low Heights}

Our results, illustrated in Figure \ref{mmaps}, present a metal-rich bulge at low heights ($|Z| \leqslant 0.50$\,kpc). 
This has been suggested by previous studies, but those were based on 2D maps ($l$, $b$), without spatial resolution 
along the line of sight. Our 3D maps show significant variations in metallicity with position within the bulge.
The side of the bulge closest to the Sun ($Y<0$) appears  metal-rich (\feh\ $\sim +0.2$), while the 
more distant parts ($Y>0$) seem to be more metal-poor (\feh\ $\lesssim -0.2$).  

Figure \ref{metvar}  collapses the information on that axis to offer a different perspective of the median metallicity as a function of 
Galactocentric distance ($R_{\rm GC} = \sqrt{X^2+Y^2}$, with $R_{\rm GC}$ set to negative values at $Y<0$). Changing
the sign of $R_{\rm GC}$ depending on whether a location is closer or further away than the Galactic center is
very useful to consider separately the more distant regions, which are more prone to systematic effects (see Section\,\ref{bias}).
The part of bulge closer to the Sun seems to be more homogeneous in metallicity, although, low metallicity regions 
are also observed. The far/distant side of the bulge exhibits in general lower metallicities, 
but those low metallicity regions can be followed, at intermediate 
Galactic longitudes ($l \sim 15\degr$), by regions of higher metallicities.

The APOGEE survey was conducted from the Northern Hemisphere, but did manage to observe 
some lower latitude regions of the Southern Galactic Hemisphere, albeit with overall poorer statistics, 
a situation now being remedied by data acquisition with the Southern Hemisphere-based spectrograph of APOGEE-2.  
In general, data from the northern Galactic latitudes appear fairly similar to those obtained from regions located south of the Galactic plane.

As discussed in previous sections, the observed variation in metallicity with heliocentric distance is suggestive of biases in the stellar sample due to
the cool limits of the model atmospheres used in the spectral analysis.
The Galactic bar can contribute to the observed variations, however, its effect should have a marked dependence on
Galactic longitude, which is not observed, and a symmetry in metallicity respect to the bar location 
would be expected, which is not apparent in the data probably because of our sample selection (See Section\,\ref{modcomp}).
The parts of the bulge closer to the Sun are the ones less affected by sample biases, and we will focus on those for the reminder of the paper.


\subsection{Vertical Gradient}

\begin{figure}
\figurenum{7}
\begin{center}
\includegraphics[trim=0cm 4.3cm 2.7cm 0cm,angle=0,scale=0.18,clip]{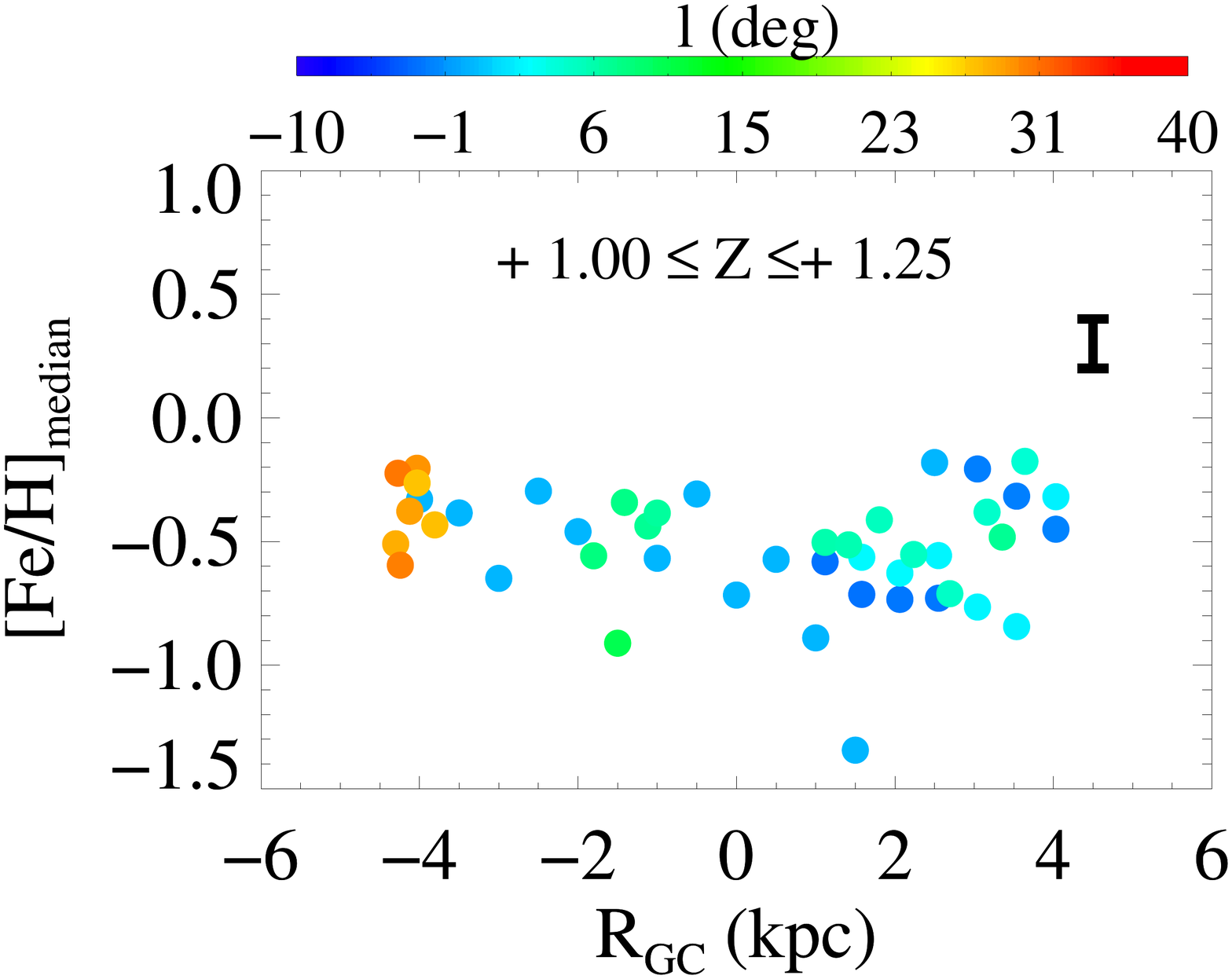}
\includegraphics[trim=5.2cm 4.3cm 2.7cm 0cm,angle=0,scale=0.18,clip]{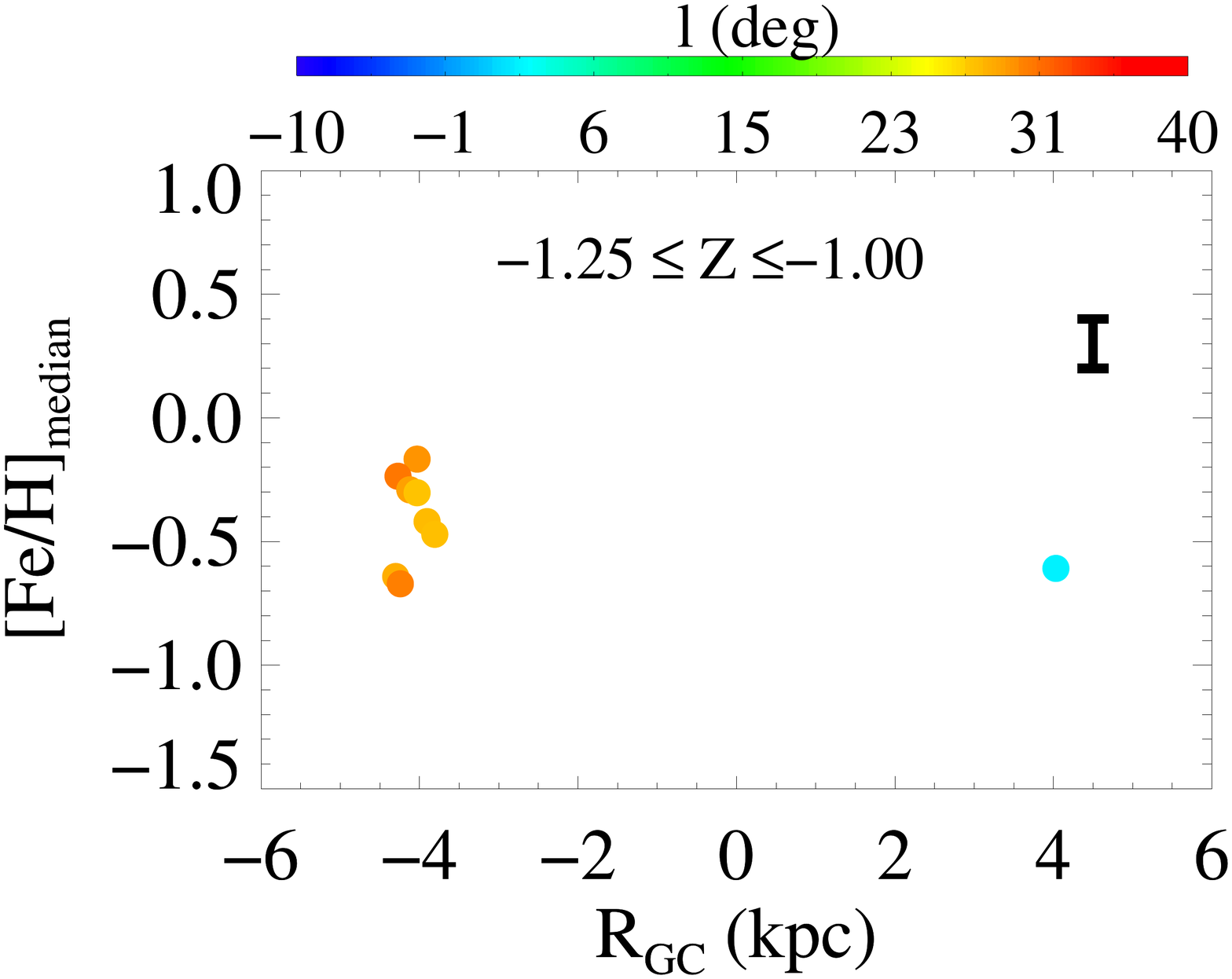}\\
\includegraphics[trim=0cm 4.3cm 2.7cm 4.00cm,angle=0,scale=0.18,clip]{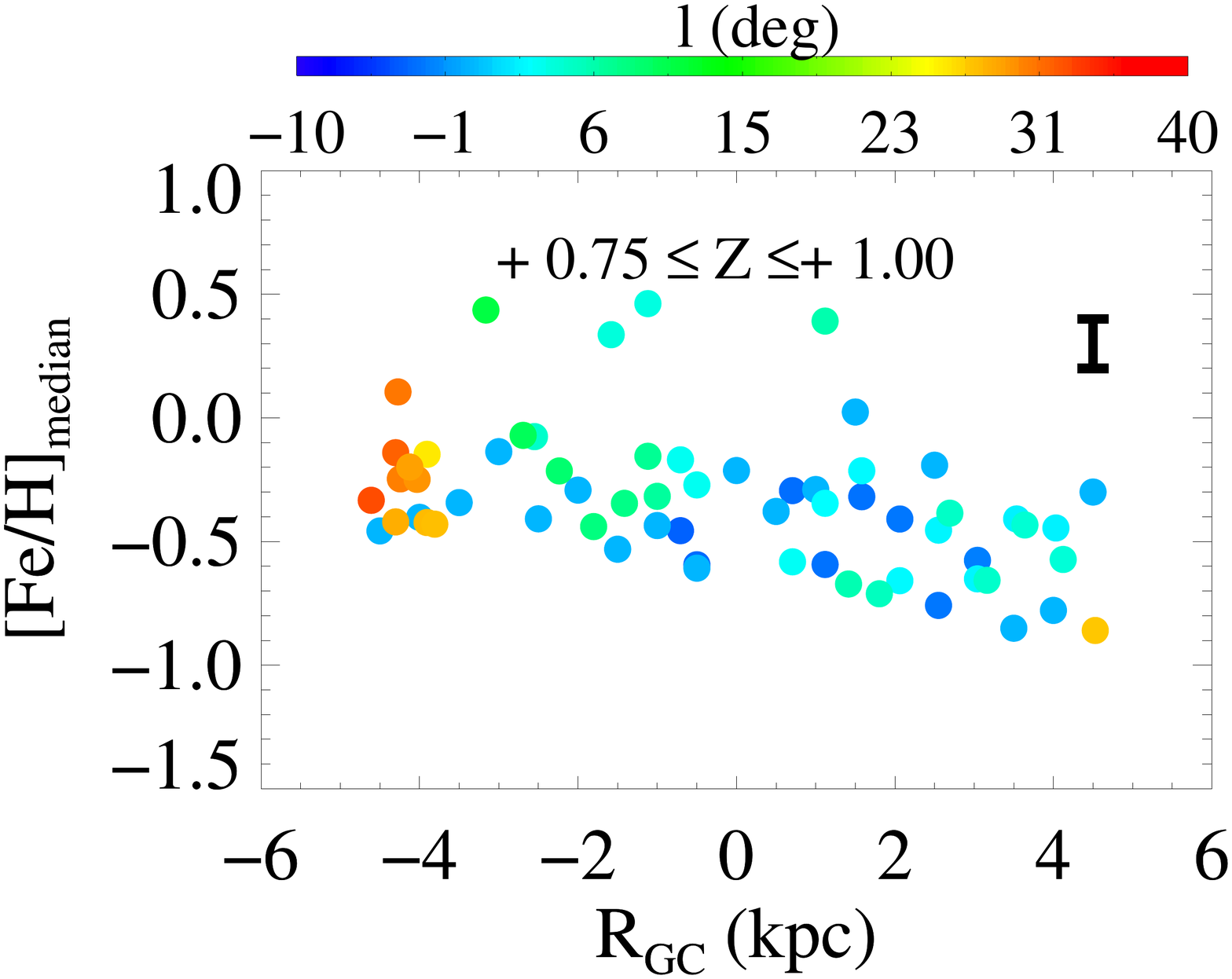}
\includegraphics[trim=5.2cm 4.3cm 2.7cm 4.00cm,angle=0,scale=0.18,clip]{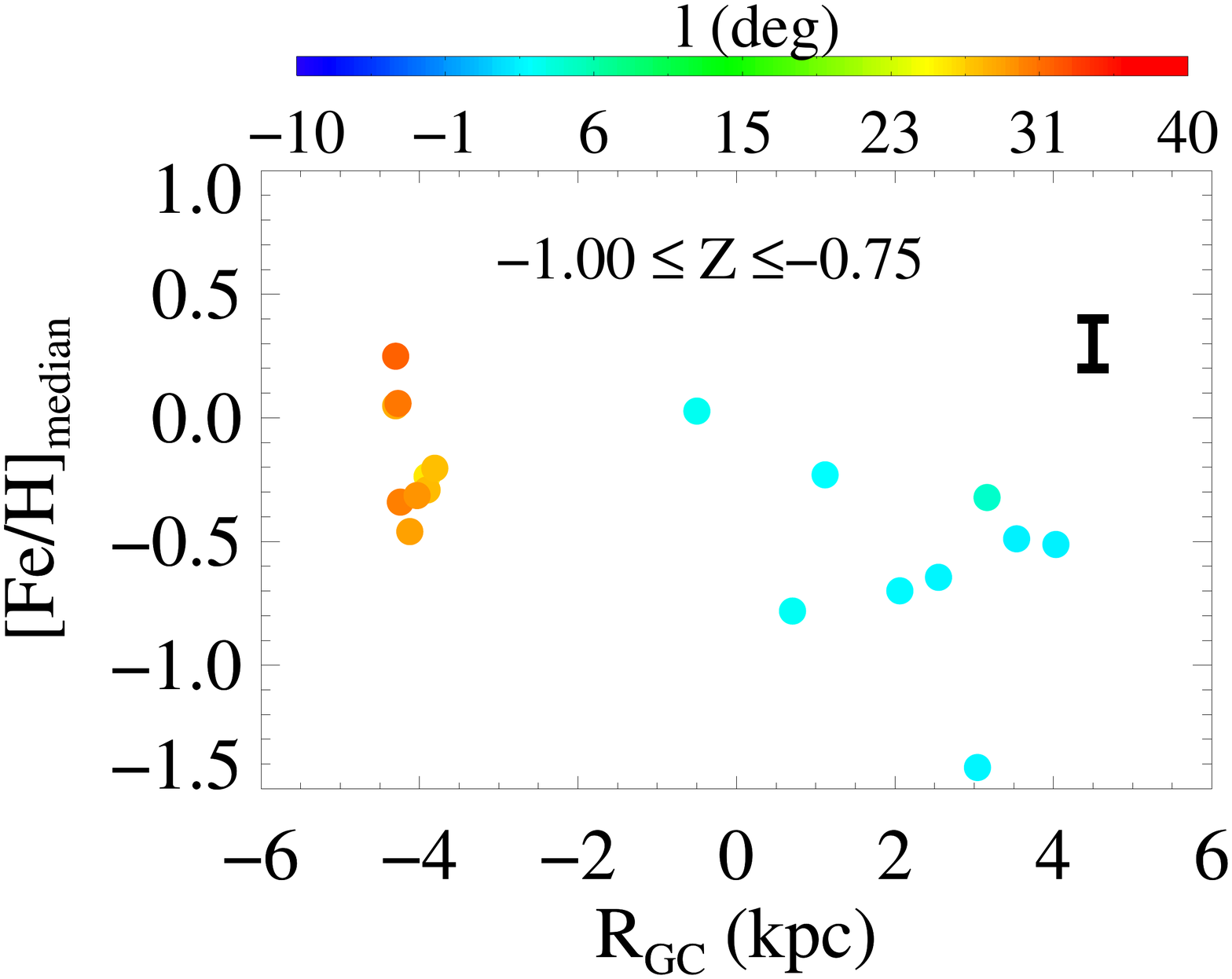}\\
\includegraphics[trim=0cm 4.3cm 2.7cm 4.00cm,angle=0,scale=0.18,clip]{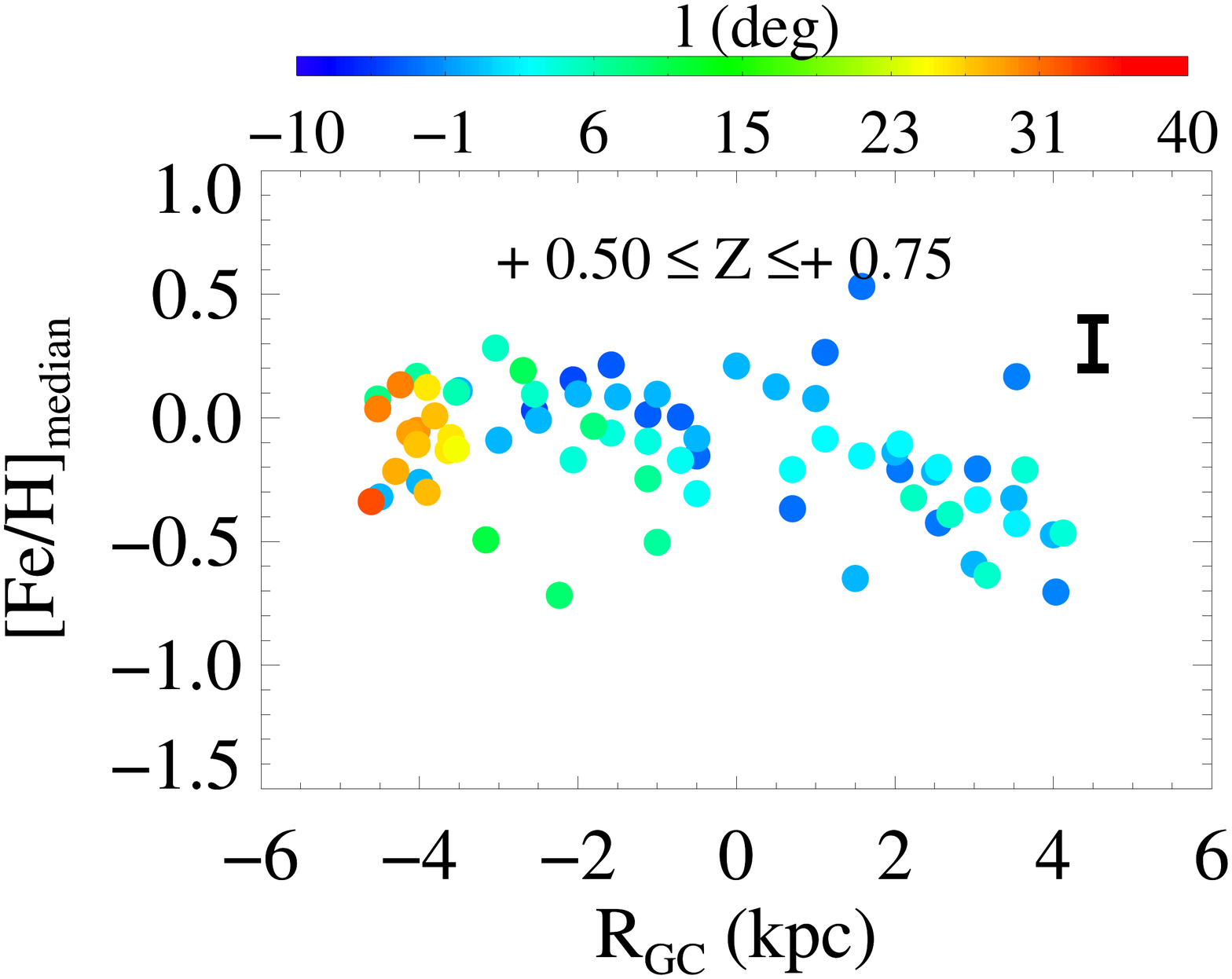}
\includegraphics[trim=5.2cm 4.3cm 2.7cm 4.00cm,angle=0,scale=0.18,clip]{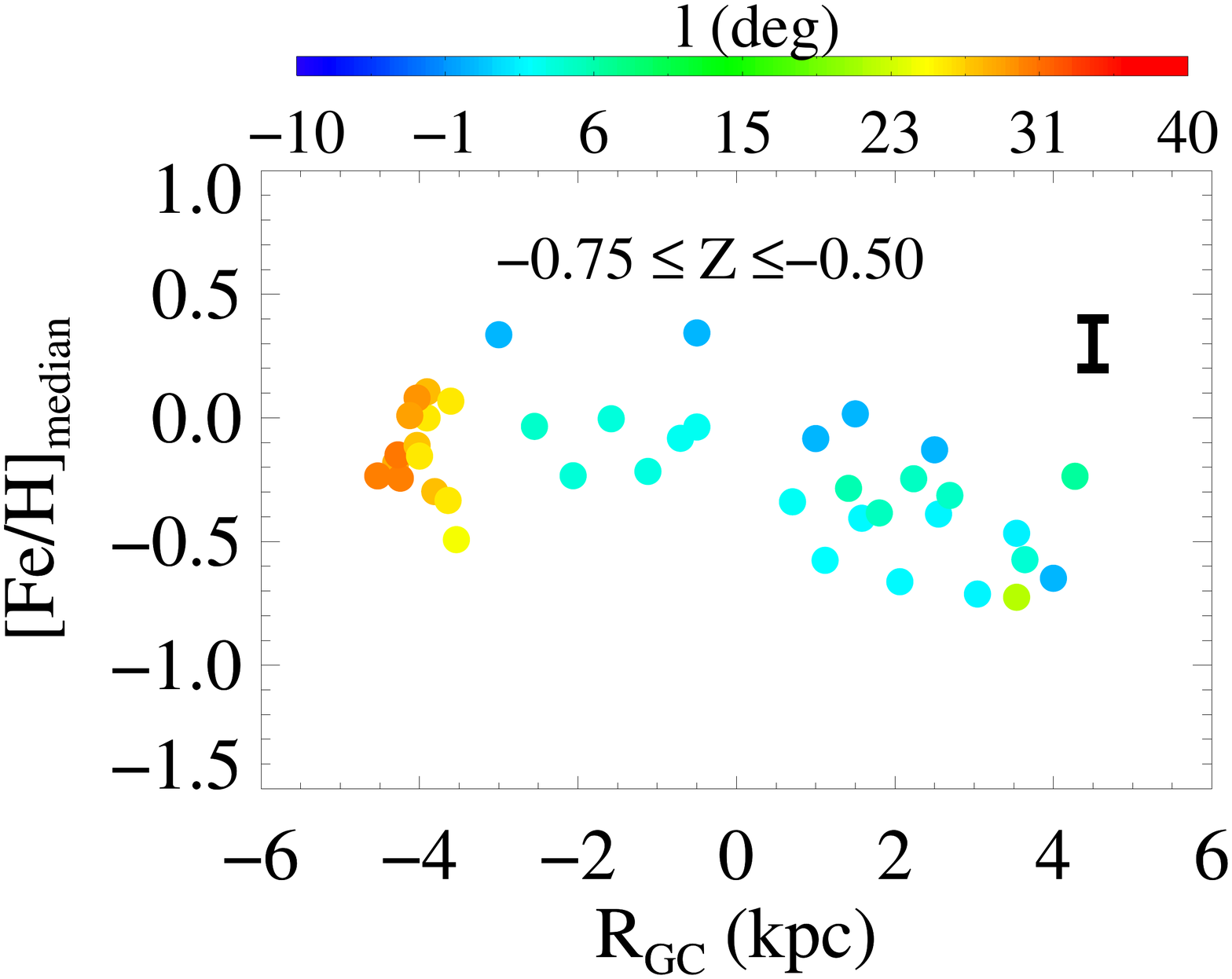}\\
\includegraphics[trim=0cm 4.3cm 2.7cm 4.00cm,angle=0,scale=0.18,clip]{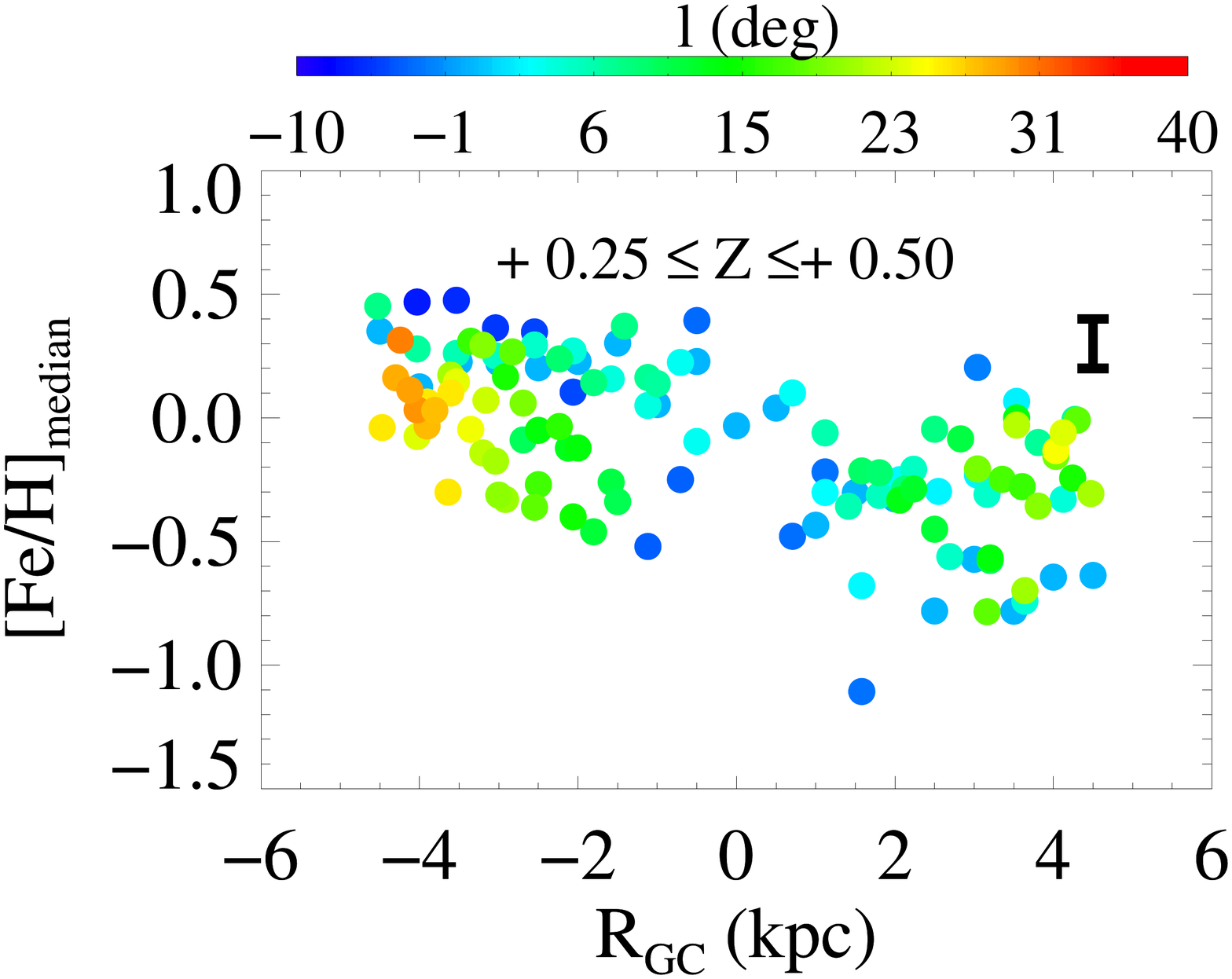}
\includegraphics[trim=5.2cm 4.3cm 2.7cm 4.00cm,angle=0,scale=0.18,clip]{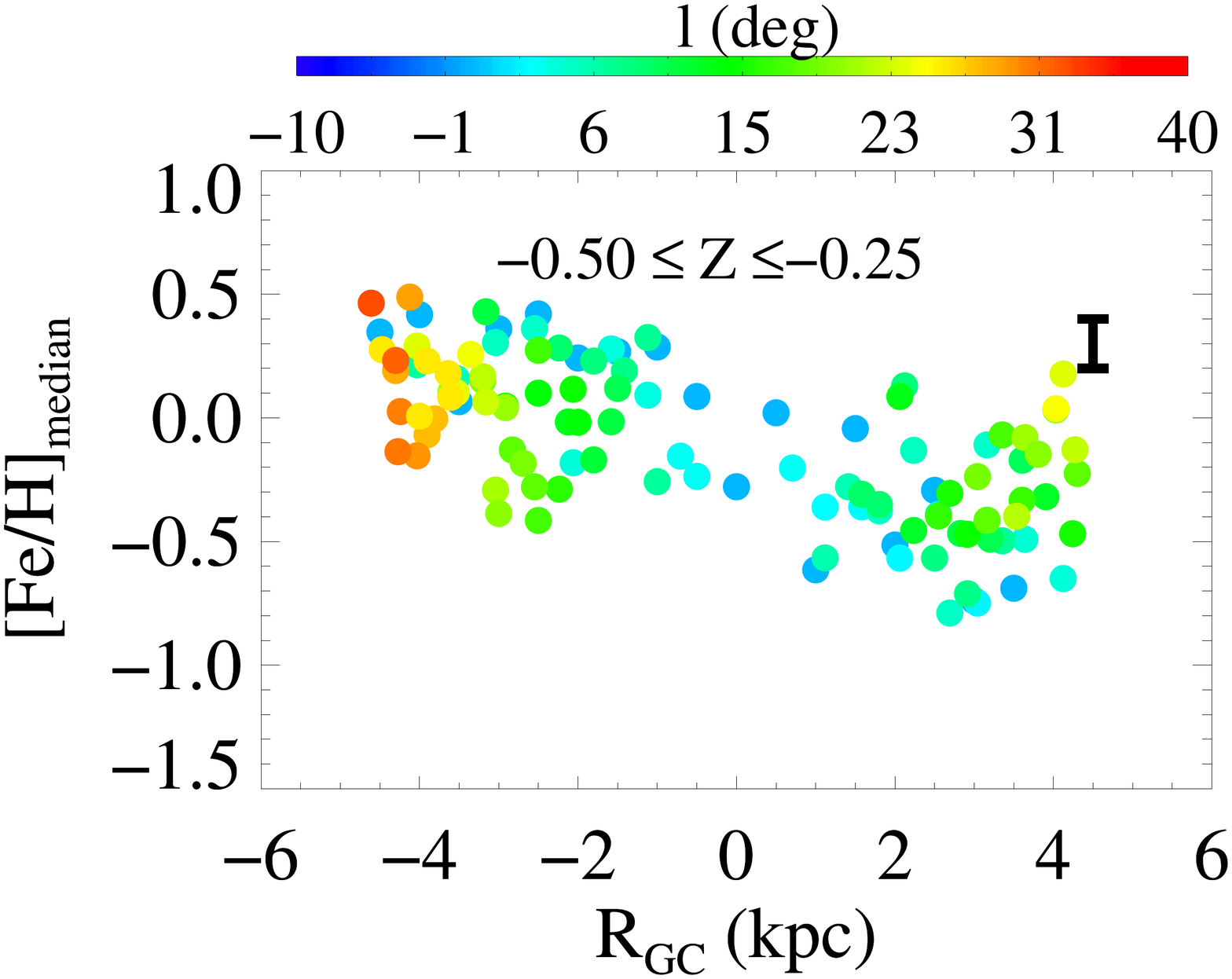}\\
\includegraphics[trim=0cm 0cm 2.7cm 4.00cm,angle=0,scale=0.18,clip]{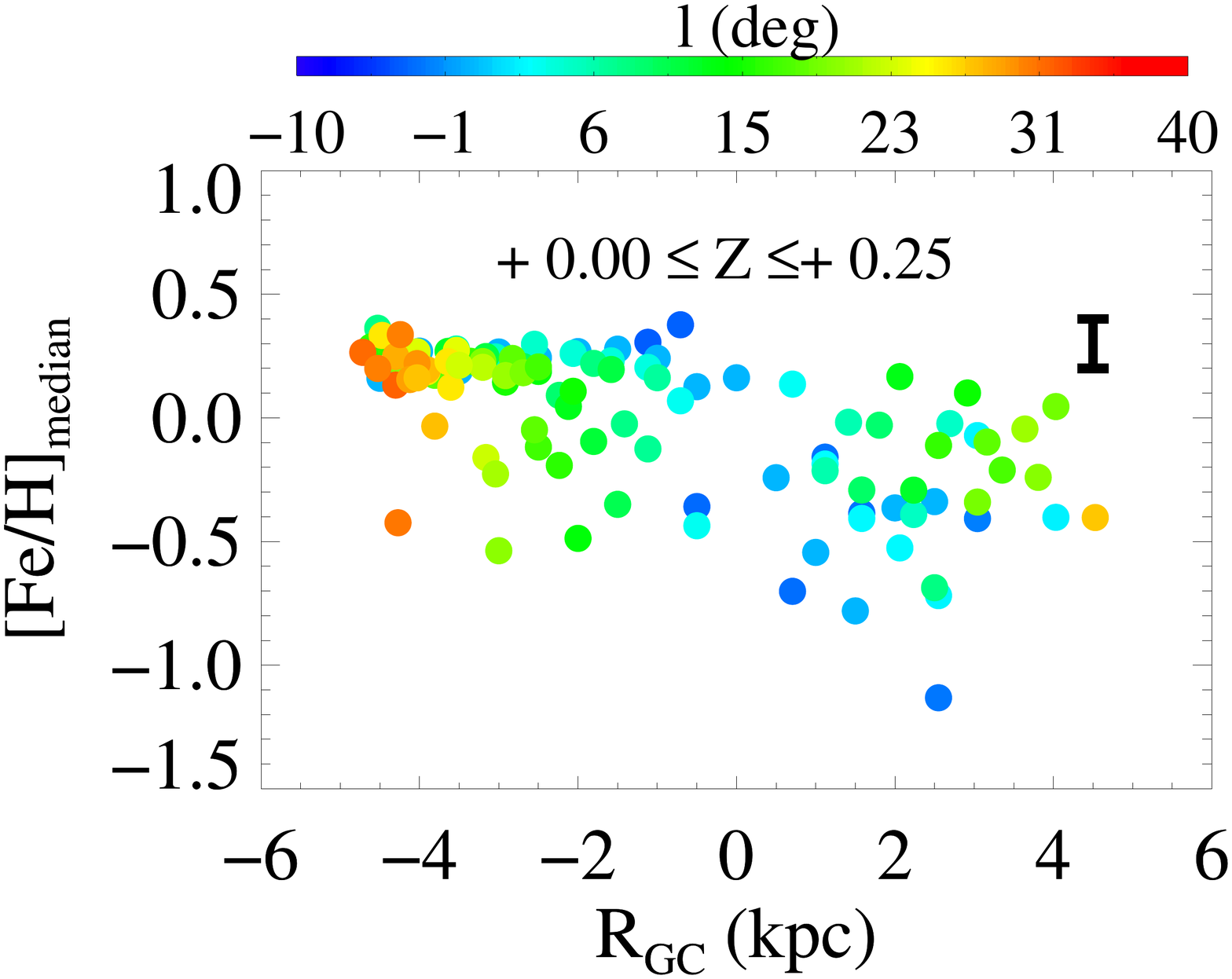}
\includegraphics[trim=5.2cm 0cm 2.7cm 4.00cm,angle=0,scale=0.18,clip]{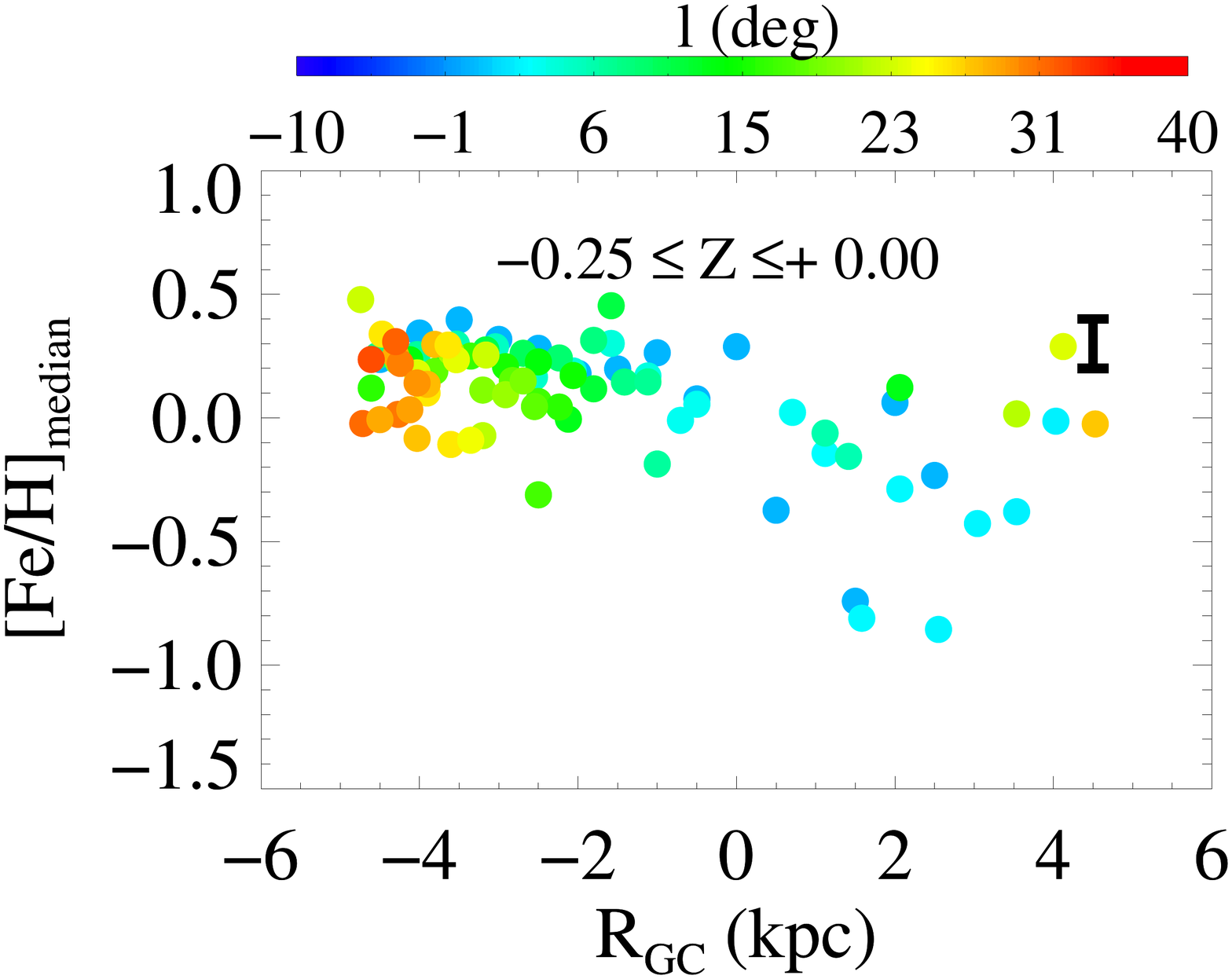}\\
\caption{Median metallicities as a function of Galactocentric distance projected in the midplane (with a typical error bar shown). Negative values of distances were adopted for the side of the Galactic Center closest to the Sun to distinguish between the near- and far-side. The color bar indicates the Galactic longitude.}
\end{center}
\label{metvar}
\end{figure}

Far from the midplane ($\abs{Z} \geqslant +0.75$\,kpc), Figure\,\ref{metvar} shows a 
more homogeneous bulge dominated by stars with relatively low metallicity (\feh\ $\lesssim -0.5$ on average). 
Interestingly, some locations show super-solar metallicity, and a couple quite low metallicities (\feh\ $<-1$). 
APOGEE's increased coverage of the low latitude bulge allows us to establish firmly the presence of a vertical  
metallicity gradient, consistent with the findings of \citet{Zoccali08}, \citet{Gonzalez13}, and \citet{Ness13}.
The gradient is no longer evident on the distant part of the bulge, but as discussed in the previous section,
our sample lacks metal-rich stars in those regions, especially closer to the plane where the extinction is stronger.

\begin{figure}
\figurenum{8}
\label{gradfig}
\begin{center}
\includegraphics[trim=2.5cm 13.0cm 2cm 1cm,angle=0,scale=0.5,clip]{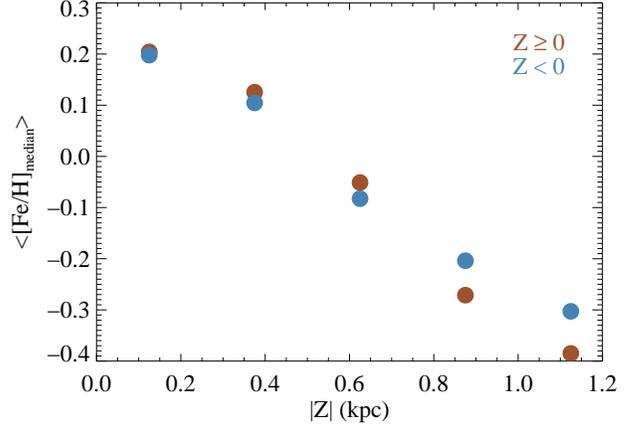}
\end{center}

\caption{Median values of the metallicities in Figure\,\ref{metvar} at different heights. Only values for regions of $-5\le R {_\mathrm{GC}\ }\mathrm{(kpc)} \le 0$ are considered.}
\end{figure}

Figure\,\ref{gradfig} shows the median metallicity in the region located in the part of the bulge closer to the Sun ($-5 \le R_{\rm GC} \le 0$)
as a function of distance from the plane. 
The slope of the vertical gradient is not constant, but it appears to be the steepest at intermediate
heights from the Galactic plane, at $+0.50 \leqslant \abs{Z}  \leqslant +1.00$\,kpc, with changes in metallicity 
of $\sim-0.2$\,dex in 0.25\,kpc. These regions have a sparser coverage in the Southern Galactic Hemisphere than in the northern one, 
however, the results for both are fully consistent.

An inner flattening of the metallicity gradient was suggested in earlier studies \citep[e.g.,][] {Ramirez00,Rich12,Babusiaux14}, 
which had data for only a few ($l$, $b$) locations. We do not only confirm the flattening, but show 
the presence of a transition region at intermediate heights, and a flattening beyond $|Z|>1$ kpc.


\begin{figure*}
\figurenum{9}
\begin{center}
\includegraphics[trim=4cm 2cm 2cm 2cm,angle=0,scale=0.8,clip]{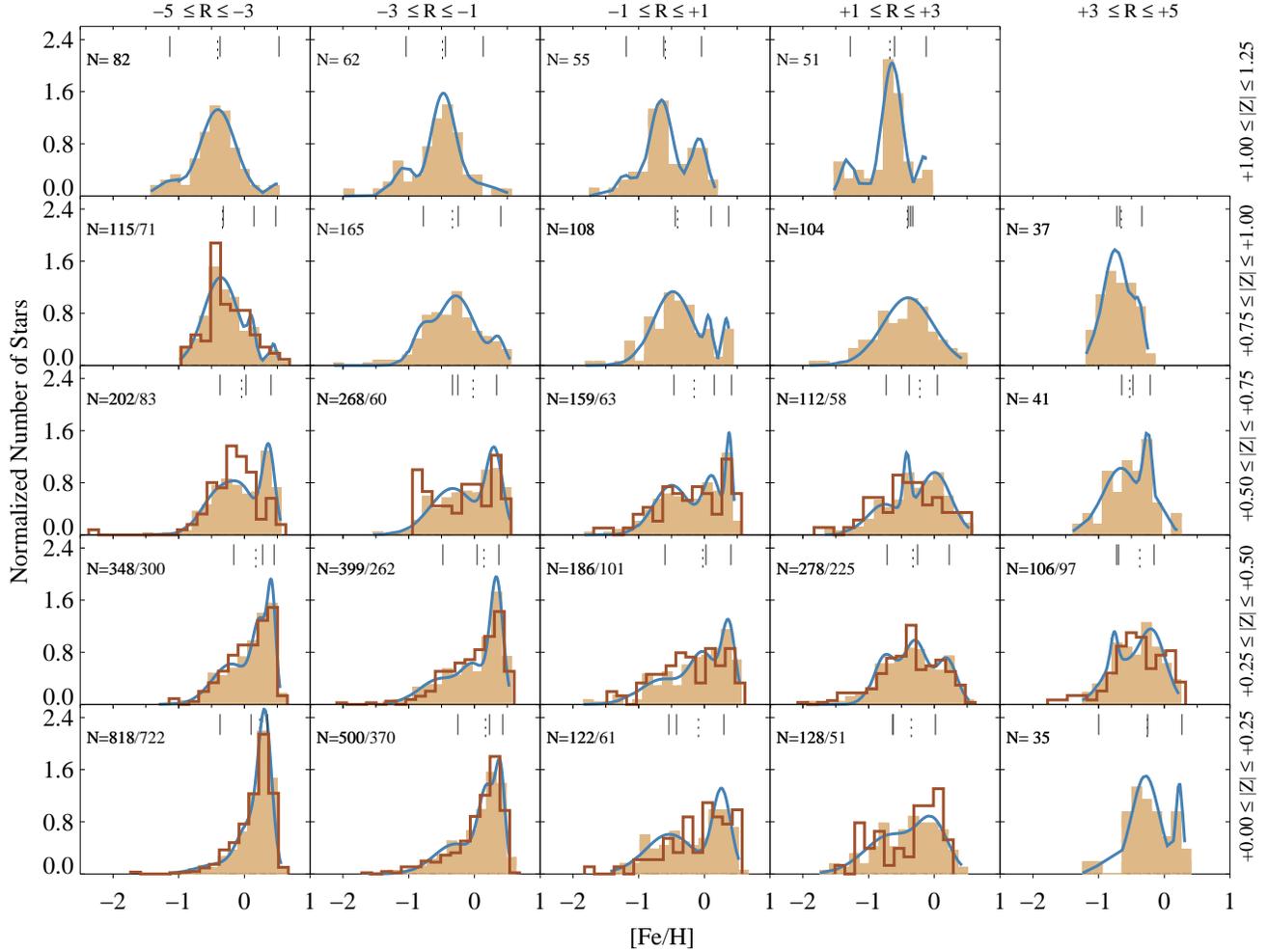}
\caption{Metallicity distribution functions ({\it tan histograms}) in bins of 0.15\,dex arranged by projected Galactocentric distance and distance from the Galactic mid-plane (including only those regions having samples of $N > 30$). Histograms are normalized to their area. 
Southern data are shown in brown and their associated number of stars is given after the slash. The three Gaussian decomposition ({\it blue curves}) is displayed only for northern regions. The vertical lines show the mean metallicity of the individual Gaussians and the median metallicity of the MDF (dashed).}
\end{center}
\label{mdffig}
\end{figure*}



\subsection{Metallicity Distribution Functions}
\label{MDFs}

There is only a few studies of the bulge MDF including low-latitude regions, and these have been 
restricted to a narrow range in ($l$, $b$) (e.g., \citealt{Rich07, Rich12, Gonzalez15, Zoccali17}). \cite{Babusiaux14} had observations in fields at $b=0\degr$, but their results were based on optical spectra acquired at a lower spectral resolution ($R=6500$). The APOGEE database now permits the most complete and accurate study of 
the distribution of individual metallicities for stars in the Galactic midplane and inner bulge.

The metallicity distributions at different projected Galactocentric radii and heights, 
in spatial bins of 2\,kpc and 0.25\,kpc respectively, are shown in Figure\,\ref{mdffig}. Only spatial 
bins including more than 30 stars are presented. Normalized MDFs are displayed in 0.15\,dex 
metallicity bins, which are twice as large as the typical metallicity uncertainty for stars in the sample. 

In the mid-plane, corresponding to the bottom row of panels in Figure\,\ref{mdffig}, 
stars from low to super-solar metallicities are observed 
(\feh$\ \sim-1\ \mathrm{to} +0.5$), but  the metal-rich stars are the dominant component.  This is particularly true
in the regions on the bulge quadrants closer to the Sun, at $R_{\rm GC}<0$, which we trust are the ones less affected
by sample biases. This metallicity range is very similar to that reported by \cite{Gonzalez15},
but it does reach significantly lower metallicities than in the studies by \cite{Rich07,Rich12}, most likely due to a smaller sample size
that makes them miss the rare very low-metallicity stars.
On the other hand, stars of lower metallicity are the major contributors far from the plane ($|Z|>0.75$).
Note the presence of a significant metal-poor contribution around the GC, 
as previously seen in APOGEE data by \citealt{Schultheis15}, and reported earlier by \cite{Babusiaux14}.

\begin{figure}
\figurenum{10}
\begin{center}
\includegraphics[angle=0,trim=4cm 5cm 8cm 0cm, scale=0.3,clip]{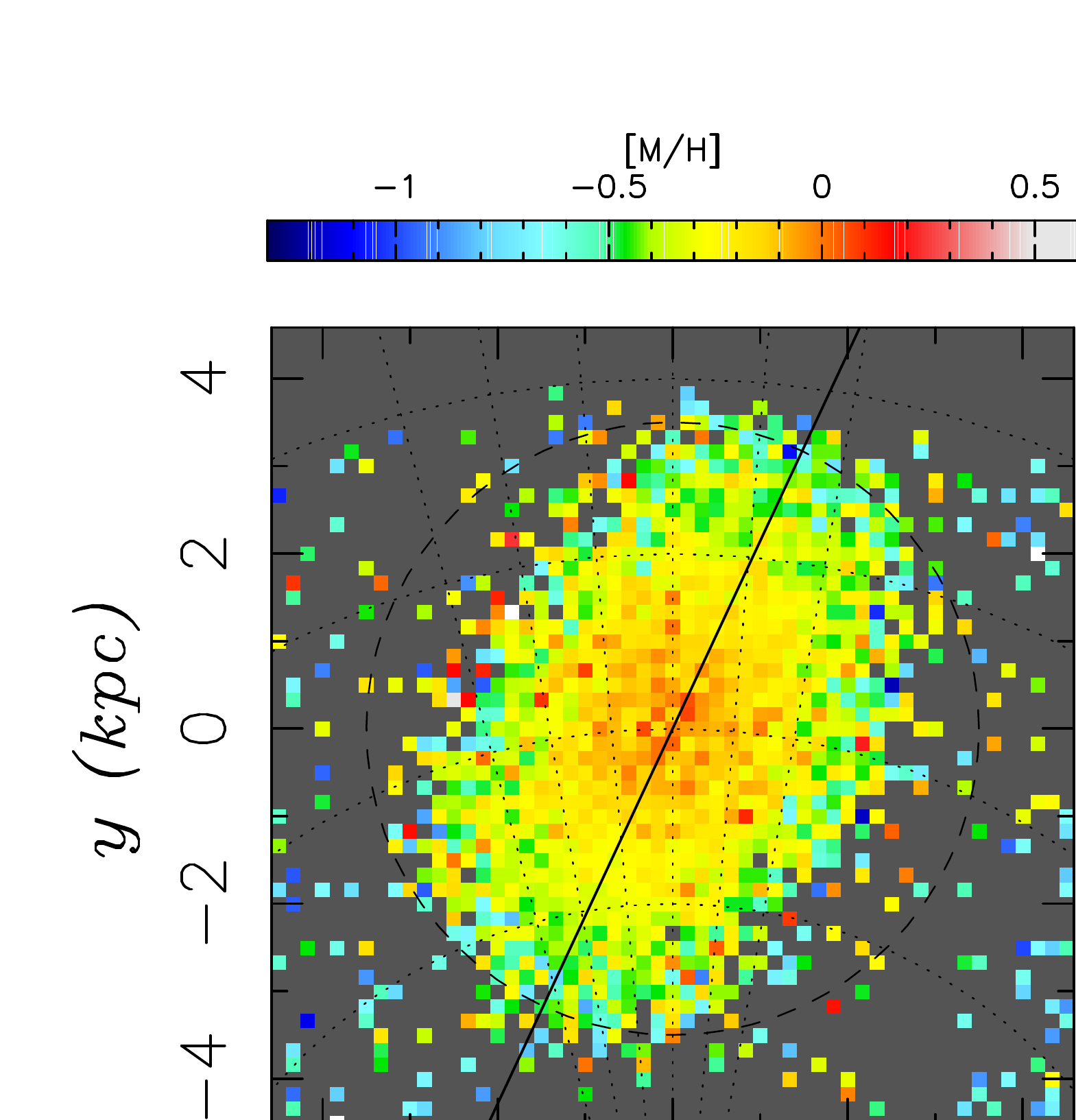}\\
\includegraphics[angle=0,trim=4cm 5.1cm 8cm 5cm, scale=0.3,clip]{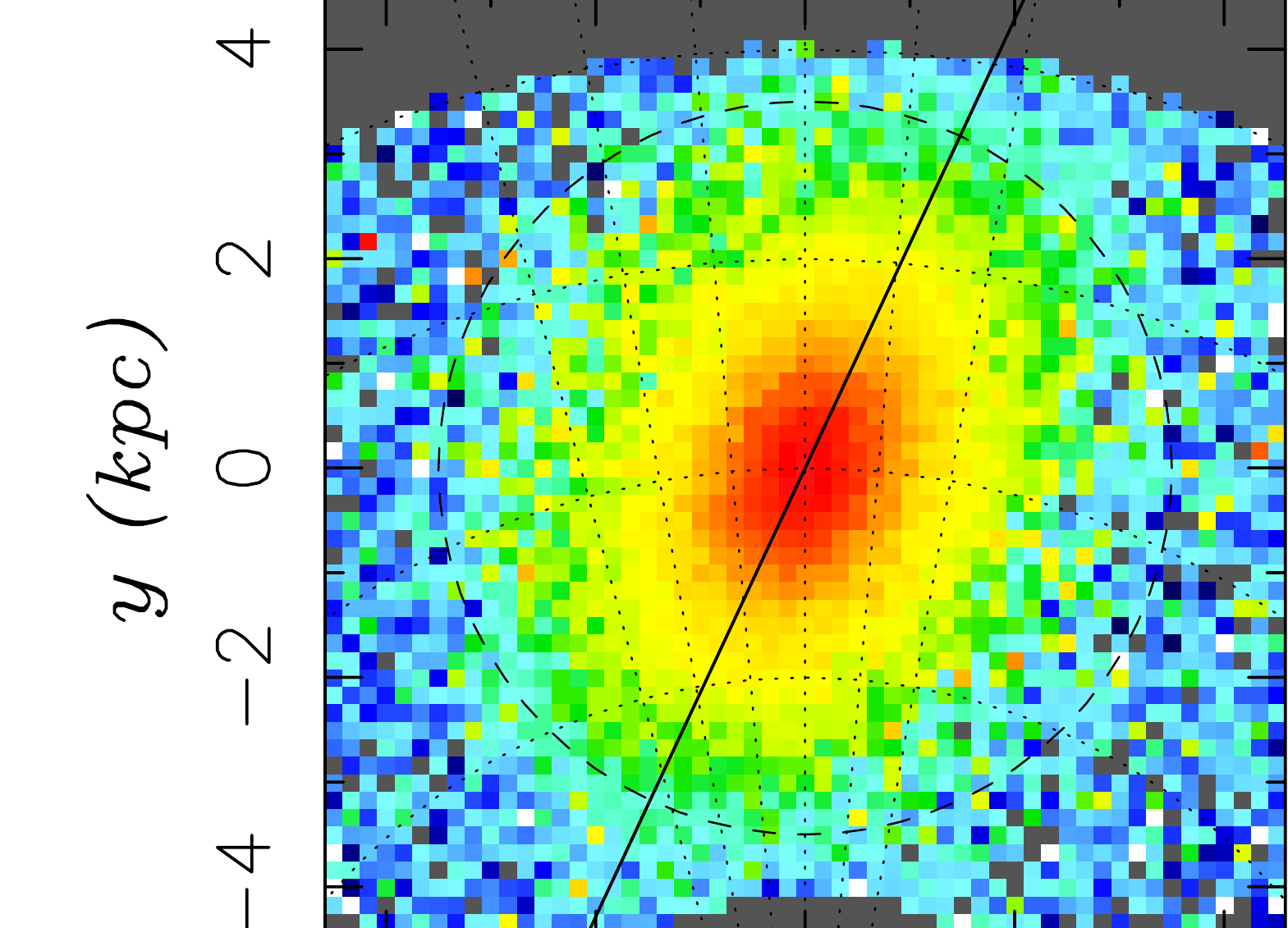}\\
\includegraphics[angle=-90,trim=5.1cm 4cm 1cm 8cm, scale=0.3,clip]{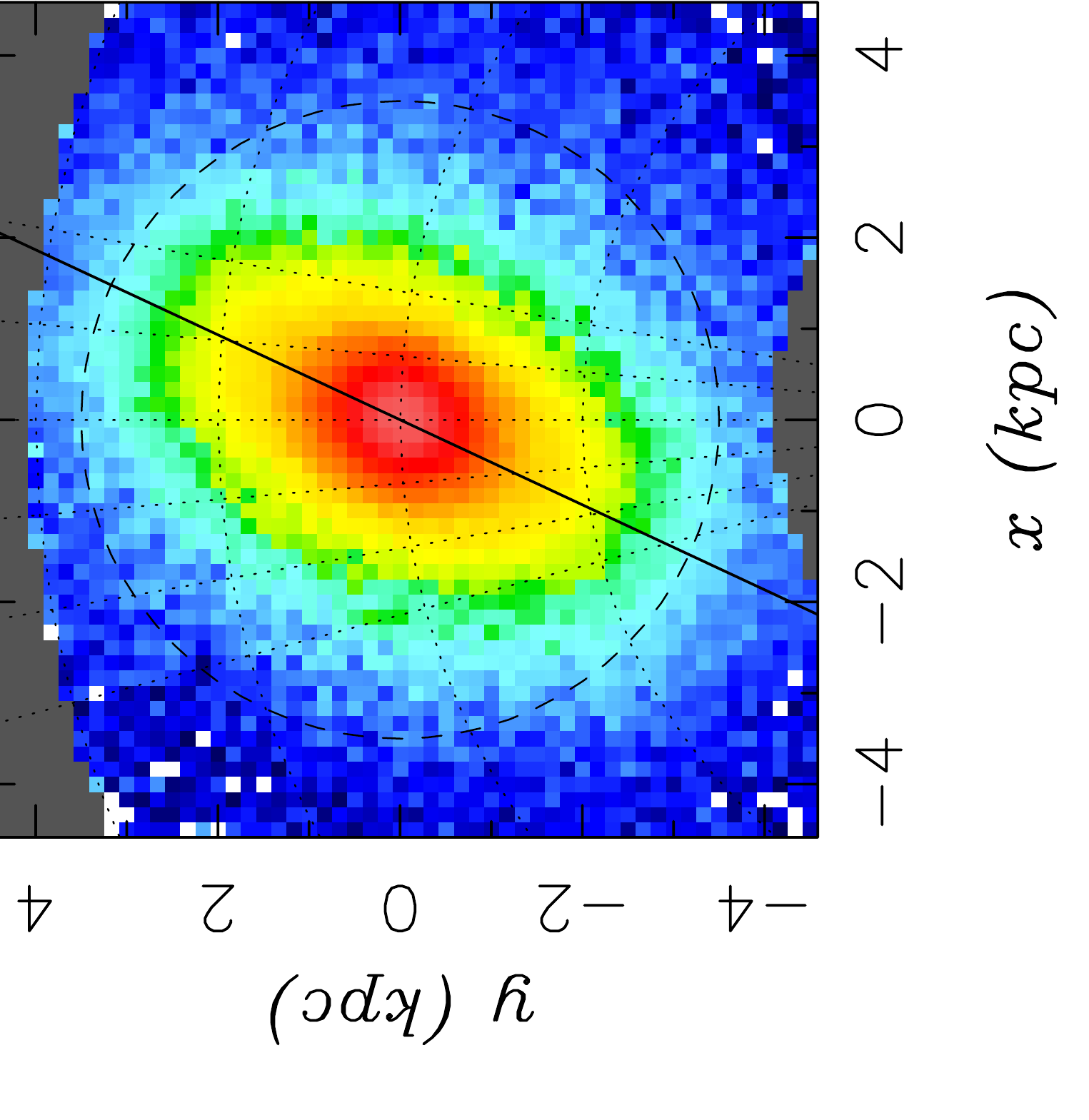}\\
%
\caption{The MVG mean metallicity distribution at $1.50\leqslant|z|\leqslant 1.00$ (top), 
$0.50\leqslant|z|\leqslant 1.00$ (middle), and $0.00\leqslant|z|\leqslant 0.50$ (bottom) for 
$4 \leqslant d \leqslant 12.0$\,kpc. Metallicities are color-coded and the dashed lines 
represent $l=0\degr, \pm5\degr$, $\pm 10\degr$, and $+15\degr$, and projected radii of 3.5\,kpc. 
The orientation of the bar is indicated by a solid line.}
\label{MVG}
\end{center}
\end{figure}



\subsection{MDF Decomposition}

The detection of different metallicity distributions in the inner Galaxy can be interpreted in 
terms of density variations in multiple overlapping metallicity components (e.g., disk, bar, classical bulge, 
and inner halo), as suggested by \cite{Ness13}, rather than a bulk change in the overall population metallicity. 
For each of the selected regions  in Figure\,\ref{mdffig},  
the distribution of the metallicities contains information about such components. 

\cite{Ness13} concluded that a minimum sample of $\sim500$ ARGOS survey stars were required 
to detect multiple metallicity components. The minimum number of stars per bin must be lower 
for the APOGEE sample, due to its greater metallicity precision: 0.05--0.09\,dex versus  0.13\,dex for ARGOS. 
\citet{Lindegren13} found that the minimum sample size required for resolving two different 
chemical distributions separated by $r$ times the standard deviation (i.e., the measurement uncertainty) could
be approximated by the expression $N_{\rm min} \simeq \exp{(0.6+13 r^{-0.8})}$. This means that two 
populations whose metallicities differ by 0.32 dex could be resolved only with a sample of $\sim$ 500 stars
measured with the precision of the ARGOS observations, while only $\sim 60$ would be required at the 
typical precision of the APOGEE bulge sample.  In our analysis we typically have more than 100-200 stars per spatial bin,
and for the most part we find a smooth variation of the distributions across neighboring regions.

The metallicity distributions  often exhibit multiple peaks, and  vary with position. 
A three Gaussian (3G) decomposition of the MDFs based on a maximum 
likelihood estimator and an analysis of jackknife samples \citep{Bovy11} returned components at four 
different metallicities, marked with vertical lines in Figure\,\ref{mdffig}: $+0.32$ (metal-rich), $+0.00$ (intermediate metallicity), and $-0.46$ and $-0.83$ (metal-poor). It should be noted that the separation of the multiple components is larger than the 
uncertainties in their fitted mean metallicities ($\sigma\leqslant 0.15$\,dex)
The four different values are related to the four distinct 
centers found from the various 3G decompositions, with generally three of these four discriminated at each location.

There is indication of a metal-poor component at \feh$\ =-1.22$ far from the plane ($\abs{Z} > +1.00$\,kpc), which
may be connected to the stellar halo \citep{Allende14}.
The fraction of very metal-poor stars detected in this study is larger than that found in the ARGOS survey: 
1.47\% versus 0.07\%, respectively, for a metallicity of \feh$\leqslant-1.5$.  
Most of the ARGOS metal-poor stars are seen at high Galactic latitude ($b\geqslant6\degr$). 
Possible explanations for the greater numbers of the current work  include a larger presence 
of this population at low heights and/or a metallicity bias.

The components peaking at [Fe/H] $\sim -0.83$ and $-0.46$, more prominent far from the plane, 
may be related to the thick disk \citep[see{\it{,}} e.g.,][]{Lee11}. For a more appropriate comparison in terms of homogeneity and proximity, compare with disk values in \citep{Hayden14}. The contribution of the component centered at \feh$=-0.46$ 
is significant in the central regions of the low bulge ($0 \leqslant \abs{Z} \leqslant +0.25$\,kpc). 
In fact, the fraction of metal-poor stars (\feh$\le-0.3$) along the midplane ranges from 7\% at the nearest location to 41\% in the GC.
Such differences are large in comparison to the noise. 

The components at metallicities  $\sim0.2 - 0.3$ resemble the distributions reported for the central 
parts of the thin disk (see{\it{,}} e.g., $R\sim5$\,kpc in Figure\,7 of \citealt{Hayden14}). This may be indicative of a bulge with a disk-origin. The solar and super-solar metallicity components have a larger contribution at low $|Z|$, 
becoming the major contributors (especially the most metal-rich) on the side closer to the Sun. 
Interestingly, the solar-metallicity component extends to 3\,kpc in radius (cylindrical coordinates). 
This component is present at low heights, independent of the set of heliocentric distance estimates 
employed (estimates other than of the current work were also investigated), however, it is not visible in the most central regions. 
More uncertain is its vertical extent, whose contribution can extends 
significantly beyond the intermediate heights depending on the adopted set.

 The values of the metallicity components found in this study ($+0.32$, $+0.00$, $ -0.46$ and $-0.83$) 
 are consistent with those previously reported in the literature. The first and third most metal-rich components 
are in good agreement with the high-spectral-resolution results from GIBS  \cite[$+0.26$ and $-0.31$;][]{Gonzalez15} and 
the Gaia-ESO survey \cite[$+0.18$ and $-0.50$;][]{Rojas14}. 
The agreement with ARGOS \cite[$+0.10$, $-0.28$, $-0.68$,  and $-1.18$;][]{Ness13} is slightly worse, possibly due to differences with their metallicity scale. Based on a common stellar sample, their estimates in the super-solar metallicity regime are lower than the APOGEE values, and that explains some of the different components identified. However, in the low-metallicity regime, there are discrepancies between the identified populations that cannot be explained by a metallicity offset. Some differences with the study of micro-lensed dwarfs \citep{Bensby17} are also observed. We note that both \cite{Ness13} and \cite{Bensby17} cover different parts of the bulge than this study. Both studies find more peaks than those obtained in our 3G decompositions, despite the smaller stellar sample in \cite{Bensby17}.
A metal-rich component has also been detected in the midplane by Babusiaux et al. (2014; $+0.20$).
The good agreement demonstrated with various literature studies as well as the small derived uncertainties 
in the metallicity decomposition offer further support for the distributions we identify in the APOGEE data. 
 

\section{Model Comparisons}
\label{modcomp}

The metallicity results of this study are compared with two different models: the 
N-body dynamical simulation of Martinez-Valpuesta \& Gerhard (2013; MVG hereafter) 
and the population synthesis model from the Besan\c{c}on Galaxy Model (Robin et al. 2012, 2014; BGM hereafter). 
The latter model relies upon more assumptions regarding the Galactic gravitational potential 
and directly aims to reproduce the observed properties of the stellar populations.

\subsection{MVG Simulations}

The MVG simulation consists of a boxy bulge that evolved from an exponential disk (Q=1.5, scale-length of 1.29\,kpc, and 
scale-height of 0.225\,kpc) embedded in a live dark matter halo and that suffered from instabilities and 
bar buckling \citep[see][]{Martinez11}. 
The resulting bar has a length of 4.5\,kpc and an orientation of 25\degr\ between the bar major axis 
and the Sun-GC axis. Metallicity was added to the simulation by assigning a radial metallicity gradient 
(\feh$=0.6-0.4R_\mathrm{GC}$; $R_\mathrm{GC}$ in kpc units) to the initial disk, which was chosen to 
reproduce the vertical metallicity variations observed by \cite{Gonzalez13}. According to the model, 
low-metallicity stars from the outer disk are mapped inwards to high latitudes producing the vertical gradients. 
The simulation snapshot we adopted was taken after the system was relaxed, $t\sim1.9$\,Gyr.

The simulation shows a metal-rich inner bulge elongated along the bar and surrounded 
by a metal-poor disk (see Figure\,\ref{MVG}) due to the initial setup. Low-metallicity stars 
come from the outer disk (no thick disk is included). The model does not show the asymmetry
in the metallicity distributions we observe between the quadrants closer to the Sun and those 
beyond the Galactic center, in line with our conclusion that those are the result of a bias in our sample.
On the other hand, we would have expected to observe the symmetry around the bar position
shown by the simulations, but we do not.

The simulation cannot reproduce the high metallicities observed in the solar neighborhood nor of the inner disk. 
Furthermore, \citet{Hayden14} show a quite flat radial gradient for the thin-disk near the bulge at $|Z| < 0.25$\,kpc. 
This is in contrast with the larger gradient adopted in MVG. Milder metallicity gradients, such as those observed near the Sun, may reproduce better  the high metallicities we observe in the midplane.

\subsection{BGM Model}

This model consists of a mixture of multiple stellar populations: bar, thin- and thick-disk, and halo. Specific 
properties are assigned as follows:

\begin{itemize}

\item  A thin disk with ages from 0 to 10 Gyr, with an age-metallicity relation from \cite{Haywood08} 
in the solar neighborhood, and a 
radial metallicity gradient of $-0.07$\,dex/kpc. Its scale-length has been constrained from a 
study of 2MASS stars counts presented in \cite{Robin12}. 

\item A bar with an age of 8 Gyr, an average solar metallicity, and no gradients. 
The shape of the bar has been determined from 2MASS color-magnitude diagrams \citep{Robin12}.

\item A thick disk having two epochs of star formation at ages of 10 and 12 Gyr. 
Its characteristics have been determined in \cite{Robin14}. 
The mean metallicities are $-0.5$ and $-0.8$, respectively, and no metallicity gradients are assumed.

\item A stellar halo, with an age of 14 Gyr, a mean metallicity of $-1.5$, and no metallicity gradient.

\end{itemize}

The kinematics for each population are computed mainly as described in \citet{Robin03} for the thin and thick disks, 
and for the stellar halo, 
as given after the updates on the age velocity dispersion relation coming from the fit to RAVE and Gaia TGAS data \citep{Robin17}. 
For the bar, the full 3D velocity field is computed using a N-body model from \citet{Debattista06}, 
scaled to fit BRAVA's data \citep{Gardner14,Robin14}.

This model has been constrained observationally,  but in that exercise no APOGEE data were used.
In the comparison below,  APOGEE data are simulated by applying the selection criteria  
introduced during the survey targeting process. The number of targets in each field are selected exactly as done 
for actual APOGEE observations. Further cuts are applied to remove regions of the \logg-vs-\teff\ 
plane compromised by  ASPCAP's limitations.
 
The sample extracted from this model is therefore restricted to $4000\leqslant$ {\teff} $(\mathrm{K}) \leqslant 
4500$. Cuts in distances are not applied to avoid introducing  uncertainties associated with the observed 
distance estimates. However, the high \teff\ cut provides a natural  culling of most of the foreground  giants.


\begin{figure}
\figurenum{11}
\begin{center}
\includegraphics[trim=0cm 5cm 0cm 0cm,angle=0,scale=0.25,clip]{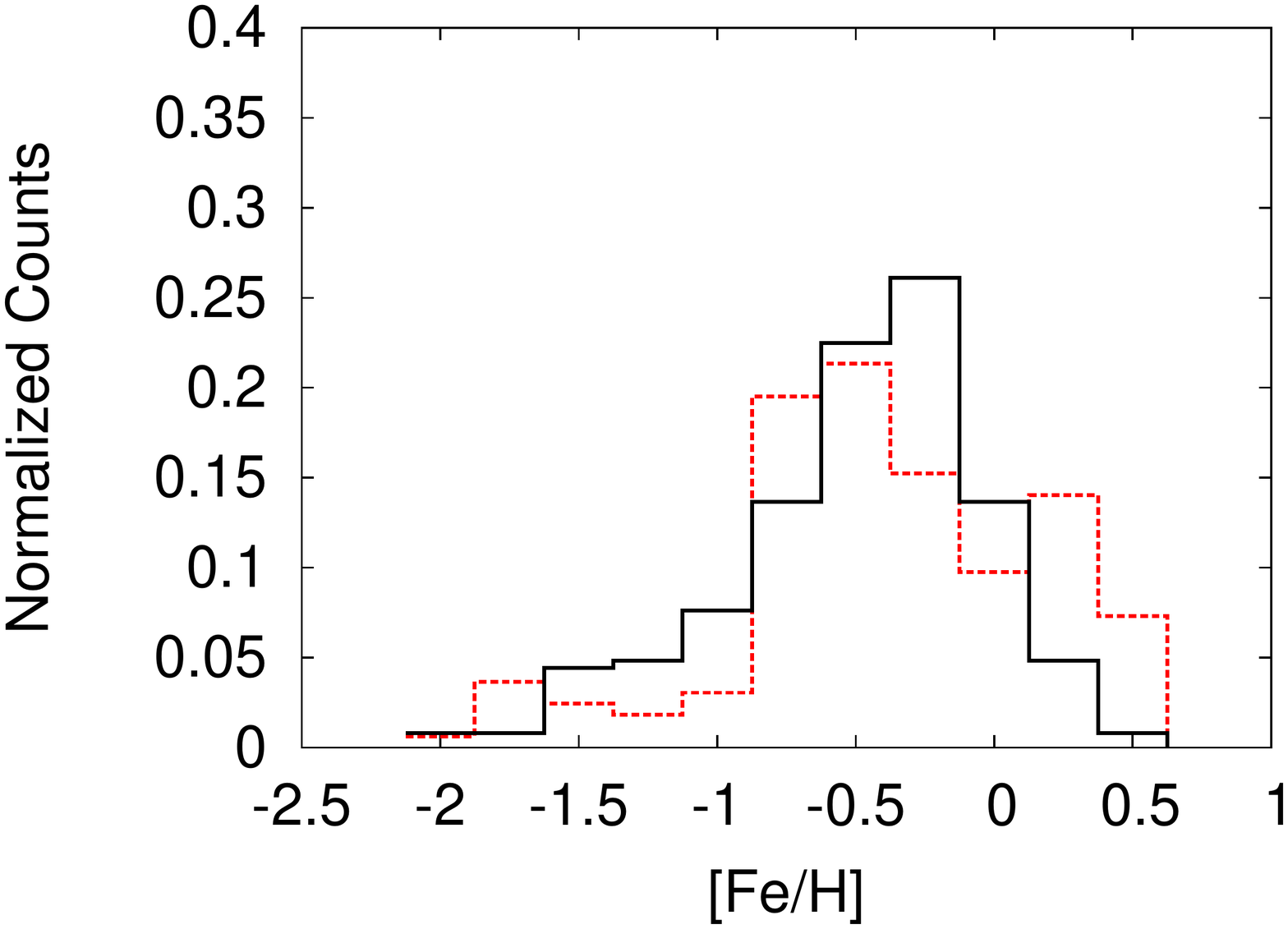} \\
\includegraphics[trim=0cm 5cm 0cm 2.2cm,angle=0,scale=0.25,clip]{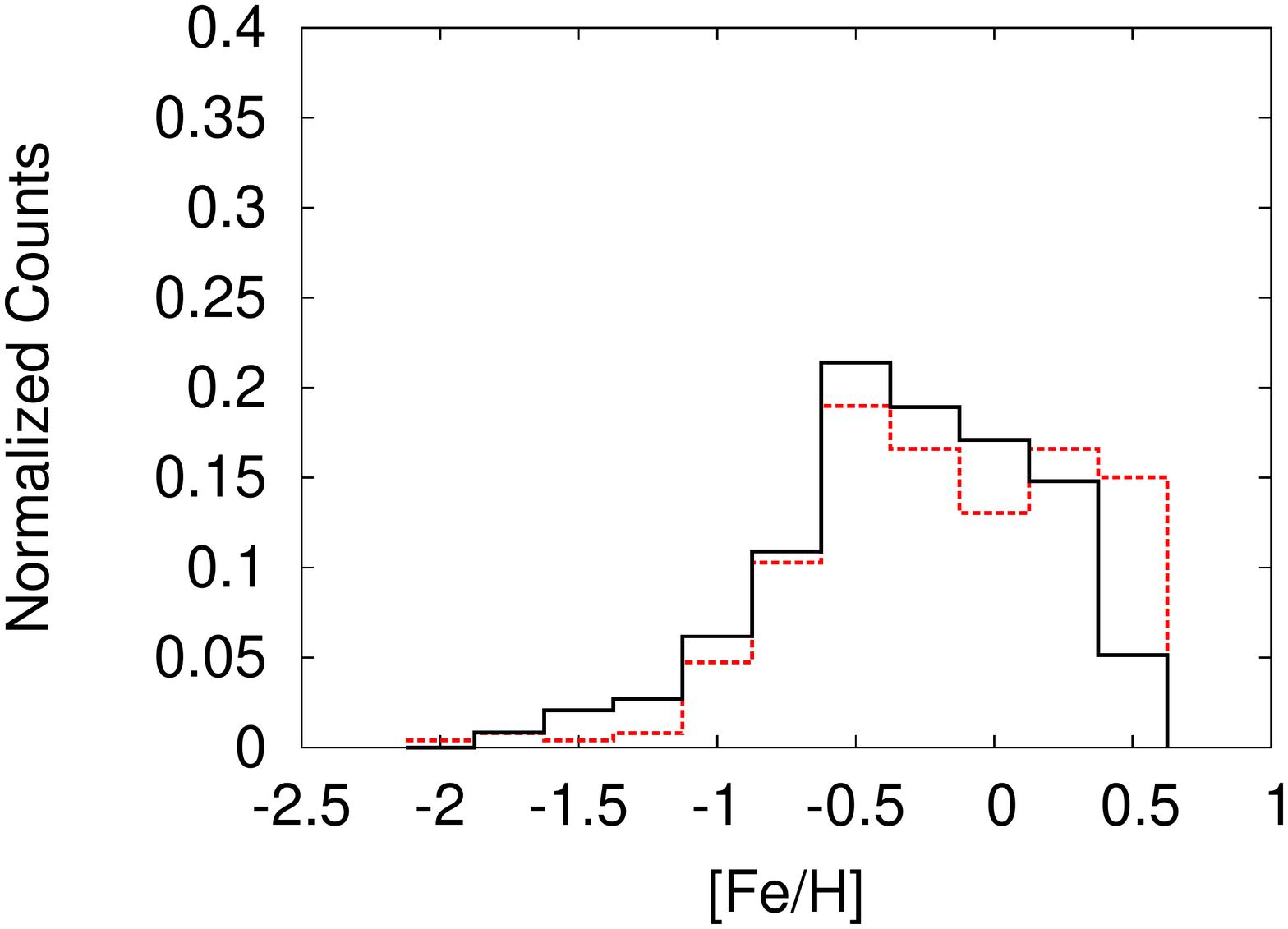}\\
\includegraphics[trim=0cm 2cm 1cm 2.2cm,angle=0,scale=0.24,clip]{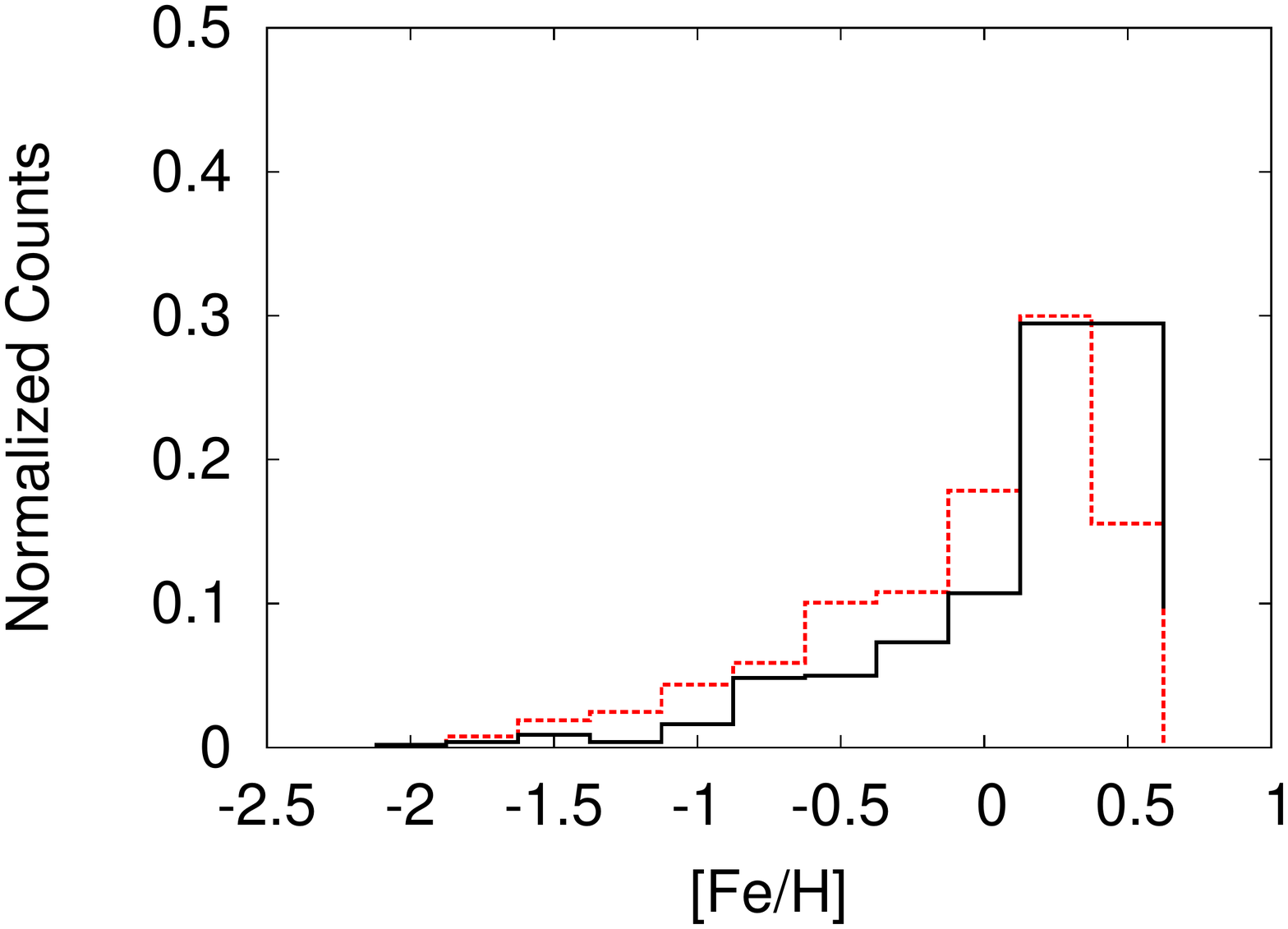}\\
\caption{Observed (black) and Besan\c{c}on simulated (red) metallicity distribution functions in bins of 0.25\,dex.  Only bulge fields with $H \leqslant 11$ and only stars with $4000\leqslant$ \teff\ (K) $ \leqslant 4500$ are considered. The MDFs are arranged by Galactic latitude. Top: $10\degr \leqslant b\leqslant 17 \degr$, middle: $4\degr\leqslant b \leqslant 10\degr$, bottom: $b\leqslant 4\degr$.}
\end{center}
\label{mdfbes}
\end{figure}


Observed and simulated MDFs are compared in three latitude bins  in Figure\,\ref{mdfbes}.
The APOGEE observations are overall well fitted by the simulations. Nonetheless, 
there are some differences, e.g., the super-solar metallicity contribution is overestimated 
in the model. The variation in metallicity  as a function of Galactic latitude is produced 
by the different proportions of the populations included in the model, distorted by the 
APOGEE selection function. In these simulations, the dominant populations are: 
the thin disk and the bar at low latitudes, the thick disk at high latitudes, and a combination of both components 
in between. There is no need to include a specific bulge component to reproduce the observed distributions.

Our observed metal-rich components and the component at $-0.46$ would be associated with 
the bar+thin-disk and the thick-disk, respectively, in the model. The inner Galaxy shows a vertical transition from metal-rich to metal-poor brought by a changeover from a region dominated by a bar+thin-disk to a thick-disk one, in line with our data. 

Still, the thick-disk would have a significant concentration in the central regions and would be the main sampled population at the far-side of the bulge. This is caused by our target and field selection, and the shortest scale-length of the thick-disk in comparison with that of the thin-disk \citep{Bensby11,Bovy12}.

Note that the chemodynamical model of \cite{Portail17} suggested that stars with metallicities as low as \feh$\sim-0.5$ are strongly barred. A boxy/peanut-like structure was also assigned to stars of that low metallicities in \cite{Ness16}. Both studies, although, are based on a metallicity grouping based on the ARGOS components. The ARGOS and APOGEE surveys are not necessarily on the same metallicity scale \citep{Schultheis17}, therefore the question is whether the findings of \cite{Ness16, Portail17}  are robust enough to the analysis assumptions; e.g., APOGEE versus their assumed ARGOS metallicity grouping.

The interpretation of the solar-metallicity component in terms of the BGM model is more challenging. In the simulation, 
the bar stops at 3.5\,kpc from the GC (the thin and thick disks extends beyond), while the 
observed solar-metallicity component extends farther. An association of this component with the bar is not straightforward, because the bar and thin-disk may not be chemically distinct. 
However, should the association be confirmed (e.g., using kinematics), our observations would 
give further support to the existence of a long bar ($\sim4.5$\,kpc), for which additional recent 
support has been offered \cite{Wegg15}. 
%

The BGM model has indications of a very metal-poor component (old thick-disk and halo) 
everywhere, but with a limited contribution at low heights and large Galactocentric radii, 
consistent with our non-detections.



\section{Conclusions}
\label{consec}

Spectroscopic observations in the IR of the central Galactic regions 
contain precious information relevant to the physical processes that participated in the formation of the bulge. That the metal-rich stars there are associated with a {\it pseudo} bulge is largely based on observations at intermediate and high Galactic 
latitudes, rather than at the latitudes typical of the bar. The nature of the metal-poor stars is somewhat more uncertain, with several proposed scenarios (e.g., a classical bulge or a thick disk).

Our study, based on high quality APOGEE data, is unique in spatial coverage, 
allowing us to carry out a thorough, in situ, investigation of the connection 
between the bulge and bar. Ours is the first large scale 3D map that combines mean metallicities and MDFs based on spectra delivering $\sigma _\mathrm{\feh} \lesssim 0.05$-$0.09$\,dex uncertainties for stars across the inner bulge. The study comprises $\sim 7545$ stars in 83 fields, largely with $|b| \leqslant 4$\degr, and over longitudes from $l=-5\degr$ to $l=32\degr$.

Stars from low to super-solar metallicities are observed in all regions.
At low- and intermediate-heights ($< 0.75$\,kpc) the APOGEE data show an overall super-solar 
metallicity bulge 
($\sim +0.2$), and a metal-poor ($\sim -0.4$) population far from the plane ($|Z|> 1.00$\,kpc) 
with a smooth transition in between. The largest vertical metallicity gradients are observed at intermediate distances from the Galactic plane,
with shallower slopes on both ends. The far-side of the bulge appears metal-poor through almost all heights, but after detailed evaluation we conclude that this effect is merely an artifact of the selection and analysis biases.

We make decompositions of the MDFs at different locations within the bulge into multiple Gaussian components, 
supported by maximum likelihood and jackknife techniques. This analysis 
suggests the presence of four metallicity components at $+0.32$ (super-solar), $+0.00$ (solar), and $-0.46$ and 
$-0.83$ (metal-poor), which are of different strength across the bulge. The two metal-rich components are observed at low and intermediate heights, but only one of them (super-solar) is observed in the most central regions. The solar component extends more than 3\,kpc in the 
 direction of the Sun, and beyond the region where we find the metal-poor components. The metal-poor component at $-0.46$, which is also centrally present at low heights, dominates at greater heights. 
 
A possible interpretation of these components, based on their metallicity and model predictions, 
is their association with the bar, the thin- and thick-disk. A comparison with the  Besan\c{c}on model indicates 
that the bar+thin-disk, and the thick-disk, contribute mostly at low- and at high $Z$-distances, respectively, 
with a smooth transition in between. Changing contributions of the different populations provide 
a simple explanation for the flattening of the vertical  metallicity gradient in the inner regions. 
Another possible interpretation (motivated by the MVG model) 
is that the bar changes the stellar orbits of the low metallicity stars
in height ($Z$) and radius, introducing chemical gradients far from the midplane. Our main discrepancy with this model is our lack of observed metal-poor regions in the midplane on the near side of the bulge, which may be indicative of an inappropriate model. Models with star formation in-situ are under construction.

The combination of chemistry and kinematics brings an improved characterization of the Milky Way central regions. 
Further progress will be possible in the near future with an expanded stellar sample from 
the ongoing APOGEE-2 survey, including observations from the Hemisphere, 
which offers a much better view of the central parts of the Galaxy. The new data and the associated 
improved statistics and coverage will be invaluable for disentangling the nature of the complex 
metallicity variations discussed in this work.

\acknowledgments

Support for A.E.G.P. was provided by SDSS-III/APOGEE. C.A.P. is grateful for support from MINECO for this research through grant AYA2014-56359-P. Sz.M. has been supported by the Premium Postdoctoral Research Program of the Hungarian Academy of Sciences, and by the Hungarian
NKFI Grants K-119517 of the Hungarian National Research, Development and Innovation Office. BGM simulations were executed on computers from the Utinam Institute of the Universit\'e de Franche-Comt\'e, supported by the R\'egion de Franche-Comt\'e and Institut des Sciences de l'Univers (INSU).

Funding for SDSS-III has been provided by the Alfred P. Sloan
Foundation, the Participating Institutions, the National Science Foundation, and the U.S. Department of Energy Office of Science. The SDSS-III website is http://www.sdss3.org/.
SDSS-III is managed by the Astrophysical Research Consortium for the Participating Institutions of the SDSS-III Collaboration including the University of Arizona, the Brazilian Participation Group, Brookhaven National Laboratory, University of Cambridge, Carnegie Mellon
University, University of Florida, the French Participation Group, the
German Participation Group, Harvard University, the Instituto de
Astrof\'{\i}sica de Canarias, the Michigan State/Notre Dame/JINA
Participation Group, Johns Hopkins University, Lawrence Berkeley
National Laboratory, Max Planck Institute for Astrophysics, Max Planck
Institute for Extraterrestrial Physics, New Mexico State University,
New York University, Ohio State University, Pennsylvania State
University, University of Portsmouth, Princeton University, the
Spanish Participation Group, University of Tokyo, University of Utah,
Vanderbilt University, University of Virginia, University of
Washington, and Yale University.

\clearpage 

\end{document}